\documentclass[aps, prd, twocolumn, amsmath, floats, floatfix, superscriptaddress, nofootinbib]{revtex4-2}
\usepackage{siunitx}
\usepackage{graphicx}
\usepackage{epsfig}
\usepackage{amsmath,amsfonts,amssymb}
\usepackage{caption}
\usepackage{subcaption}
\usepackage{wrapfig}

\usepackage{verbatim}
\usepackage{float}
\usepackage{morefloats}
\usepackage[dvipsnames]{xcolor}
\usepackage{booktabs}
\usepackage[normalem]{ulem}
\usepackage{multirow}
\usepackage{makecell}
\usepackage{tabulary}
\usepackage{cancel}
\usepackage{xcolor}

\usepackage{blindtext}      
\graphicspath{{./figures/}}   
\usepackage{hyperref}       
\hypersetup{
	colorlinks   = true, 
	urlcolor     = blue, 
	linkcolor    = blue, 
	citecolor    = red,  
}
\usepackage{booktabs}       
\usepackage{multirow}       
\usepackage{adjustbox}      
\usepackage{cellspace}
\usepackage{tablefootnote}  
\usepackage{physics}                           
\newcommand{\matx}[1]{\boldsymbol{#1}}       
\usepackage{enumitem}
\setlist[itemize]{noitemsep, topsep=0cm, leftmargin = 1.2cm}        
\setlist[enumerate]{noitemsep, topsep=0cm, leftmargin = 1.2cm}      

\usepackage{ragged2e} 
\usepackage{subcaption} 
\DeclareCaptionJustification{justified}{\justifying} 

\usepackage[capitalise]{cleveref}
\usepackage{comment}
\crefname{section}{Sec.}{Secs.}
\crefname{table}{Tab.}{Tabs.}
\crefname{figure}{Fig.}{Figs.}
\crefname{equation}{Eq.}{Eqs.}
\crefname{appendix}{Appendix\ }{Appendix\ }

\definecolor{bostonuniversityred}{rgb}{0.8, 0.0, 0.0}

\begin{document}

\title{\boldmath Convolutional Neural Networks for the classification of glitches in gravitational-wave data streams \unboldmath}

\author{Tiago~Fernandes}
\email{tiagosfernandes@ua.pt}
\affiliation{Departamento de F\'{i}sica da Universidade de Aveiro, Campus de Santiago, 3810-183 Aveiro, Portugal}

\author{Samuel~Vieira}
\email{samuel.vieira@ua.pt}
\affiliation{Departamento de F\'{i}sica da Universidade de Aveiro, Campus de Santiago, 3810-183 Aveiro, Portugal}

\author{Antonio~Onofre}
\email{antonio.onofre@cern.ch}
\affiliation{Centro de F\'{\i}sica das Universidades do Minho e do Porto (CF-UM-UP), Universidade do Minho, 4710-057 Braga, Portugal}

\author{Juan~Calder\'on~Bustillo}
\email{juan.calderon.bustillo@gmail.com}
\affiliation{Instituto Galego de F\'{i}sica de Altas Enerx\'{i}as, Universidade de
Santiago de Compostela, 15782 Santiago de Compostela, Galicia, Spain}
\affiliation{Department of Physics, The Chinese University of Hong Kong, Shatin, N.T., Hong Kong}

\author{Alejandro \surname{Torres-Forn\'e}}
\email{alejandro.torres@uv.es}
\affiliation{Departamento de Astronom\'{\i}a y Astrof\'{\i}sica, Universitat de València, C/ Dr Moliner 50, 46100, Burjassot (València), Spain}
\affiliation{Observatorio Astron\'omico, Universitat de València, C/ Catedrático Jos\'e Beltr\'an 2, 46980, Paterna (València), Spain}

\author{Jos\'e A.~Font}
\email{j.antonio.font@uv.es}
\affiliation{Departamento de Astronom\'{\i}a y Astrof\'{\i}sica, Universitat de València, C/ Dr Moliner 50, 46100, Burjassot (València), Spain}
\affiliation{Observatorio Astron\'omico, Universitat de València, C/ Catedrático Jos\'e Beltr\'an 2, 46980, Paterna (València), Spain}

\begin{abstract}
We investigate the use of Convolutional Neural Networks (including the modern ConvNeXt network family) to classify transient noise signals (i.e.~glitches) and gravitational waves in data from the Advanced LIGO detectors. First, we use models with a supervised learning approach, both trained from scratch using the Gravity Spy dataset and employing transfer learning by fine-tuning pre-trained models in this dataset. 
Second, we also explore a self-supervised approach, pre-training models with automatically generated pseudo-labels. Our findings are very close to existing results for the same dataset, 
reaching values for the F1 score of 97.18\% (94.15\%) for the best supervised (self-supervised) model. We further test the models using actual gravitational-wave signals from LIGO-Virgo's O3 run. Although trained using data from previous runs (O1 and O2), the models show good performance, in particular when using transfer learning. We find that transfer learning improves the scores without the need for any training on real signals apart from the less than 50 chirp examples from hardware injections present in the Gravity Spy dataset. 
This motivates the use of transfer learning not only for glitch classification but also for signal classification.
\end{abstract}

\maketitle

\section{Introduction}
\label{section: introduction}

The era of Gravitational-Wave (GW) Astronomy started in 2015 with the detection of the GWs emitted in the coalescence of two black holes~\cite{LIGOScientific:2016aoc} by the Advanced LIGO detectors \cite{LIGOScientific:2014pky}. To date, the LIGO-Virgo-KAGRA (LVK) collaboration \cite{LIGOScientific:2014pky,VIRGO:2014yos,KAGRA} has reported results from three observing runs (O1, O2, and O3) which comprise 90 confident detections. All signals observed are consistent with being produced in compact binary coalescences (CBCs), namely mergers of binary black holes (BBH), binary neutron stars (BNS), and binaries of a neutron star and a black hole 
(NSBH)~\cite{LIGOScientific:2017vwq,LIGOScientific:2018mvr,LIGOScientific:2020ibl,LIGOScientific:2021usb,LIGOScientific:2021djp}. 
With the start of the fourth observing run (O4) in May 2023, the number of confident detections is expected to significantly increase, yielding an estimated CBC rate of about one detection per day~\cite{gw_detection_rate}. 

GW detectors operate in extremely low noise conditions which may be disrupted by the ground motion surrounding the detectors or earthquakes, storms or even anthropogenic sources of noise~\cite{LIGOScientific:2019hgc}. 
More than 200 thousand auxiliary channels are constantly monitored to minimize instrumental noise~\cite{LIGOScientific:2016gtq}. These include angular drift of optics, light transmitted through mirrors as well as actuation signals used to control optic position in order to ensure optical cavity resonance~\cite{LIGOScientific:2019hgc,LIGOScientific:2016gtq}. With a higher detection rate, transient noises of instrumental or environmental origin, commonly dubbed ``glitches"~\cite{Zevin2017a}, will become increasingly more of a concern for the LVK detectors. Since glitches can mimic actual GW signals they hinder the sensitivity of the detectors by increasing the false-alarm rate of true GW signals. Reducing the rate of glitches requires a deep understanding of their nature and physical origin, which in some cases is unclear. Blip glitches constitute a particularly harmful example for the LIGO detectors, where an average rate of two such glitches per hour of data was measured during O1 and O2~\cite{Cabero:2019}. 

Many different approaches to classify noise transients and subtract them from strain data have been developed over the years~\cite{Allen:2005,Biswas:2013,Cornish:2015,Powell:2015,Mukund:2017,Powell:2017,George2018,Llorens-Monteagudo:2018ubm,Razzano:2018,Nitz:2018,Davis:2019,Miquel:2019,Torres-Forne:2020eax,Ormiston:2020,Colgan:2020,Chatziioannou:2021,Merritt:2021,Davis:2022,Ding:2022,Razzano:2023}. 
Current LVK efforts include e.g.~\texttt{BayesWave}~\cite{Cornish:2015} and \texttt{gwsubtract}~\cite{Davis:2019}, which model glitches using sine-Gaussian wavelets and use a linear subtraction algorithm  to remove them from the data stream. 
Moreover, GW searches (e.g. \cite{Usman:2015kfa,Messick:2016aqy,Chandra2022}) also use strategies to veto glitches, as e.g.~the $\chi^2$ time-frequency discriminator of~\cite{Allen:2005} or a sine-Gaussian veto~\cite{Nitz:2018}.
Glitch classification and mitigation using probabilistic principal component analysis has been put forward in~\cite{Powell:2015,Powell:2017,Merritt:2021}. Moreover, the {Gravity Spy} citizen-science project~\cite{Zevin2017a} has allowed to confidently divide Advanced LIGO glitches into classes, and provides a reliably classified glitch catalog that can be used by noise classification and subtraction methods. 
Glitches from the Advanced Virgo detector have also been introduced in more recent versions of Gravity Spy. A similar citizen-science project focused on the Advanced Virgo detector, {GwitchHunters}~\cite{Razzano:2023}, introduced the localization of glitches in time-frequency space, as well as the search for correlations in the detector's auxiliary channels. 
A variety of methods that are becoming increasingly important for glitch characterization are Machine Learning (ML) methods. Those have already been extensively applied to this topic (see~\cite{Cuoco:2020} and references therein). 
For instance, ML and Deep Learning (DL) methods have been used to predict the existence of glitches  using only auxiliary channel data~\cite{Biswas:2013,Ormiston:2020} or using both the GW strain and auxiliary channels~\cite{Colgan:2020}. 
Furthermore, dictionary learning techniques have been used to reconstruct and remove blips from GW data streams in~\cite{Miquel:2019,Torres-Forne:2020eax}. 

Another interesting line of work frames the glitch classification problem as a computer vision task, where glitches are converted from a time-series to spectrogram images, suitable for classification taking advantage of the rapid advances of DL methods for computer vision. 
In Ref.~\cite{Razzano:2018} this approach was successfully tested with simulated glitches, using a custom-made Convolutional Neural Network (CNN). 
After this initial success, images from the Gravity Spy dataset have also been used to train CNNs for glitch classification. First, custom-made CNNs with two to five layers were trained from scratch for 200 epochs, and individual models achieved satisfactory results, with accuracies over 95\%~\cite{Bahaadini2018}. 
Subsequently, deeper pre-trained networks with more powerful architectures were trained on the same dataset, and it was found that transfer learning reduced training time and improved results, reaching the state-of-the-art accuracy of 98.8\% in the Gravity Spy dataset~\cite{George2018}. In that work, popular CNN architectures like Inception~\cite{Ioffe2015, Szegedy2016}, ResNets~\cite{He2016}, and VGGs~\cite{Simonyan2015} were trained using transfer learning, for up to 100 epochs, and it was found that, for some architectures, transfer learning allowed to achieve accuracies higher than 98\% within 10 epochs.

Recent advances in computer vision research, like the discovery of better training recipes~\cite{Smith2018, Narang2018a, Wightman2021, Loshchilov2019} or more powerful architectures~\cite{Liu2022a, Dosovitskiy2020, Liu2021, tan2021}, could potentially improve glitch classification even more. 
With this in mind, this paper focuses on the classification of glitches in image format (spectrograms), using the latest available techniques to attempt to improve the state-of-the-art. 
In particular, we aim at improving present-day results by using the 1cycle training policy~\cite{Smith2018}, trying a more modern and powerful CNN architecture family, the ConvNeXt~\cite{Liu2022a}, training models using both supervised and self-supervised methods, and using the typical three subset split (train/validation/test) for better assessment of the model-generalization performance. We also analyze in this paper the use of transfer learning applied to O3 data. It should be stressed that the networks are trained using glitches from the Gravity Spy catalog, which only includes O1 and O2 data, and they are not retrained or fine-tuned on O3 data. 
We find that transfer learning significantly increases the flexibility of the algorithms and their performance. 
This approach might be advantageous for future GW observing runs, as it is not as sensitive to changes in the detector background noise.



This paper is organized as follows:  Section~\ref{section: dl methods} discusses the supervised and self-supervised DL methods that we employ for our study. In Section~\ref{section: dataset} we present the dataset used. Section~\ref{section: results} contains our results on glitch classification including, as well, a preliminary analysis of the performance of our methods with O3 data and actual GW signals. Finally, in Section~\ref{section: conclusion} we present our main conclusions and outline our plans for future work.

\section{Deep Learning methods}
\label{section: dl methods}

Deep Learning is a subfield of Machine Learning that relies on artificial neural networks with many intermediate layers, each of which builds on the representations learned by the previous layer in order to automatically develop increasingly complex and useful representations of the data~\cite{Lecun2015}.
While shallow learning methods, i.e. methods with very few layers, rely on carefully handcrafted features to produce useful results, DL requires no feature engineering, as it can autonomously find what are the useful features, considerably simplifying ML workflows.

According to the experience\footnote{Here, ``experience" is used with same meaning as in Tom Mitchell's Machine Learning definition~\cite{mitchell1997}.} the algorithms acquire during the training phase, DL methods can be categorized as unsupervised, supervised or self-supervised algorithms. 
In all cases, the algorithms experience an entire dataset, $\matx{X}$, which is a collection of $m$ data points $\vb*{x}^{(i)}=(x_0, x_1, ..., x_n)$, each of which has $n$ features. 
Unlike unsupervised algorithms, in supervised learning algorithms each example $\vb*{x}^{(i)}$ has, in addition, an associated label, $y^{(i)}$, which is the target for the prediction of the point. Similarly, self-supervised learning uses the same dataset composed of $n$ samples, with targets for the predictions indicated by the labels. However, the labels of those same samples are replaced by automatically generated labels, referred to as pseudo-labels. In this work, both supervised and self-supervised algorithms are used.

\subsection{Supervised Deep Learning}

The ResNet architecture family \cite{He2016} is used in this investigation as a baseline for the classification task. ResNets are a family of CNNs \cite{LeCun1989, Lecun1998} which use residual connections, i.e.~direct paths between non-adjacent layers, to allow training deeper models.
In the original implementation, ResNets end with a fully connected layer with 1000 units, corresponding to the number of classes in the ImageNet dataset~\cite{Fei-Fei2010}. For other applications, the number of neurons in this last layer is changed to match the number of classes.

The models were trained using the \texttt{fastai} library~\cite{Howard2020a}, taking advantage of its implementation of the 1cycle training policy~\cite{Smith2018}, and training using the AdamW optimizer~\cite{Loshchilov2019}. In addition, the learning rates were selected using the learning-rate finder~\cite{Smith2018}. Mixed precision training \cite{Narang2018a} was also used to reduce the memory requirements and speed up training.

In this work we employ the cross-entropy loss, which for the case of a single example, is given by
\begin{equation}
    \label{eq: cross entropy loss}
    \mathcal{L}(\vb*{\theta}) = - \sum_{k=1}^K {y}_k \log (\hat{p}_k).
\end{equation}
Above, ${y}_k$ denotes the target probability of class $k$, typically either equal to 0 or 1, $K$ denotes the total number of classes (equal to the number of neurons in the output layer), and $\hat{p}_k$ denotes the probability that the example belongs to class $k$.
A weighted loss function was also tried, to penalize more the model for mistakes in the less represented classes, by introducing a class-dependent weight, $w_k$, in the cross-entropy loss:
\begin{equation}
    \label{eq: cross entropy reweighted}
    \mathcal{L}(\vb*{\theta}) = - \sum_{k=1}^K w_k ~ {y}_k \log (\hat{p}_k).
\end{equation}
This approach was tried using class weights inversely proportional to the number of samples in each class,
\begin{equation}
    w_k = \frac{1}{N_k},
\end{equation}
with $N_k$ being the total number of samples of class $k$. 
In addition, the Effective Number of Samples (ENS) approach~\cite{Cui2019}, which obtains the weights associated with each class using 
\begin{equation}
    w_k = \frac{1 - \beta}{1 - \beta^{N_k}},
\end{equation}
was also tried. Here, 
$\beta$ is a hyperparameter that allows to control the re-weighting degree. Its value was fixed to $\beta = 0.99$, following the suggestion in~\cite{Cui2019}. 
In both cases, the class weights are normalized by dividing each class weight by the sum of the class weights and multiplying by the number of classes, in order to create weights closer to unity. 

Moreover, the use of the focal loss function, introduced in Ref.~\cite{Lin2020}, was also tested, since it adds to the cross-entropy loss a modulating factor which down-weights easy examples, that is, examples where the model is more confident. The focal loss is 
defined by
\begin{equation}
    \label{eq: focal loss}
    \mathcal{L}(\vb*{\theta}) = - \sum_{k=1}^K w_k ~ (1 - \hat{p}_k)^\gamma ~ y_k \log (\hat{p}_k),
\end{equation}
where $\gamma$ is the focusing parameter, which was kept to the default value of 2.0~\cite{Lin2020}.

In addition to training ResNets from scratch, transfer learning~\cite{Zhuang2021} has also been considered in our investigation. Transfer learning is a technique that allows to take advantage of knowledge learned in one domain, typically from a task that has many labelled examples, and to apply it to a different but related task which may have a limited number of examples. This technique is one of the cornerstones of the \texttt{fastai} library, as it allows to train more accurate models more quickly, and using less data, as compared to training the same models from scratch \cite{howard2020}. We found it interesting to explore transfer learning here to also understand if the networks (pre-trained in datasets which are different from the ones they are applied to) improve performance (in comparison to when no transfer learning is used), also allowing them to be more robust against changes in the detector background.

In the default \texttt{fastai} transfer-learning approach, after assigning the pretrained weights, the final layer is replaced with a fully connected layer with the correct number of outputs, and it is initialized with entirely random weights. Then, the networks are trained with \texttt{fastai}'s \texttt{fine\_tune} training schedule \cite{fastai_finetune}, which first trains the last layer with the others kept frozen, and then trains the whole network using lower learning rates, while also taking advantage of discriminative learning rates.

For the networks trained with transfer learning, the ConvNeXt network family~\cite{Liu2022a} was also tried. The design of this CNN family is inspired on the successful Vision Transformers such as ViT~\cite{Dosovitskiy2020}, allowing it to achieve very competitive performances (see e.g.~\cite{fine-tune}).
More details about the architectures used in this work can be found in the appendix of \cite{fernandes2022}.

\begin{figure}[!t]
    \centering
    \includegraphics[scale=0.26]{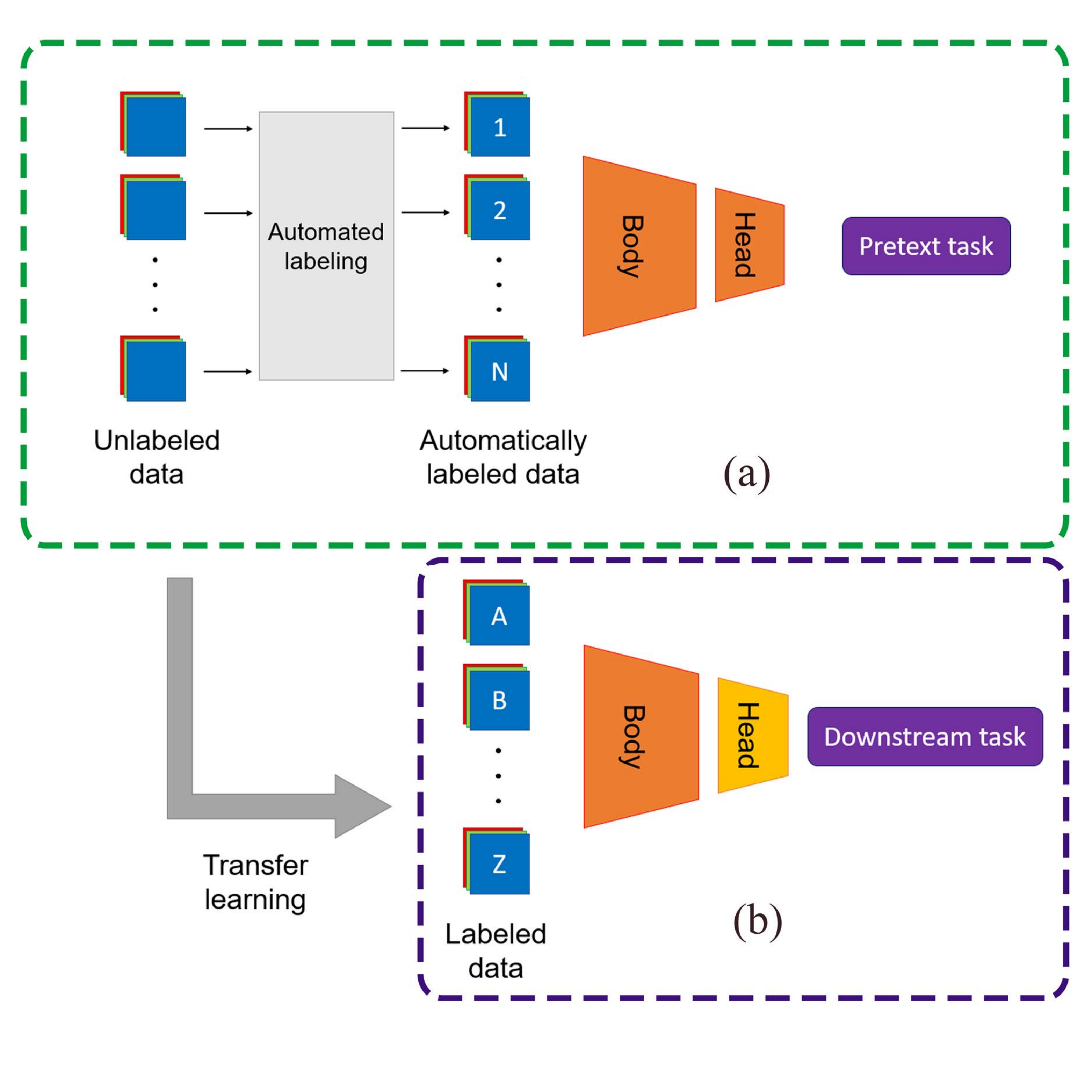}
    \caption{Scheme of the self-supervised DL framework during training. In the dashed green box (a), labels are automatically generated for a pretext task (i.e.~the image transformation imposed on the data is to be predicted by the model). In the dashed purple box (b), the architecture's last layer is changed to solve a related problem using labeled data.}
    \label{fig:ssl}
\end{figure}

\subsection{Self-supervised Deep Learning}
\label{subsection: methods self-sup}

Self-supervised DL is a recently introduced method in the field of ML~\cite{ssl_survey_2019}. This method involves training a model for a pretext task before training the model for the desired (downstream) task, as sketched in Figure~\ref{fig:ssl}. 
For instance, a model can be trained to detect what kind of visual transformation is applied to an image and, after that, a transfer of knowledge be done to another model that is trained with labeled data for the downstream task. 
To solve the pretext task, the samples contained in the dataset must have their pseudo-labels generated, as observed in box (a) of Figure~\ref{fig:ssl}. In this case, each one is associated with the type of visual transformation applied to the image. The model is then trained, allowing it to learn features/patterns about the images in the context of this so-called pretext task. The training section is approached in the same way as in supervised learning, with the objective of minimizing the loss function in order to match the model's prediction to the targets, in this case given by the pseudo-labels.

\begin{figure*}[t]
    \centering
    \includegraphics[scale=0.47]{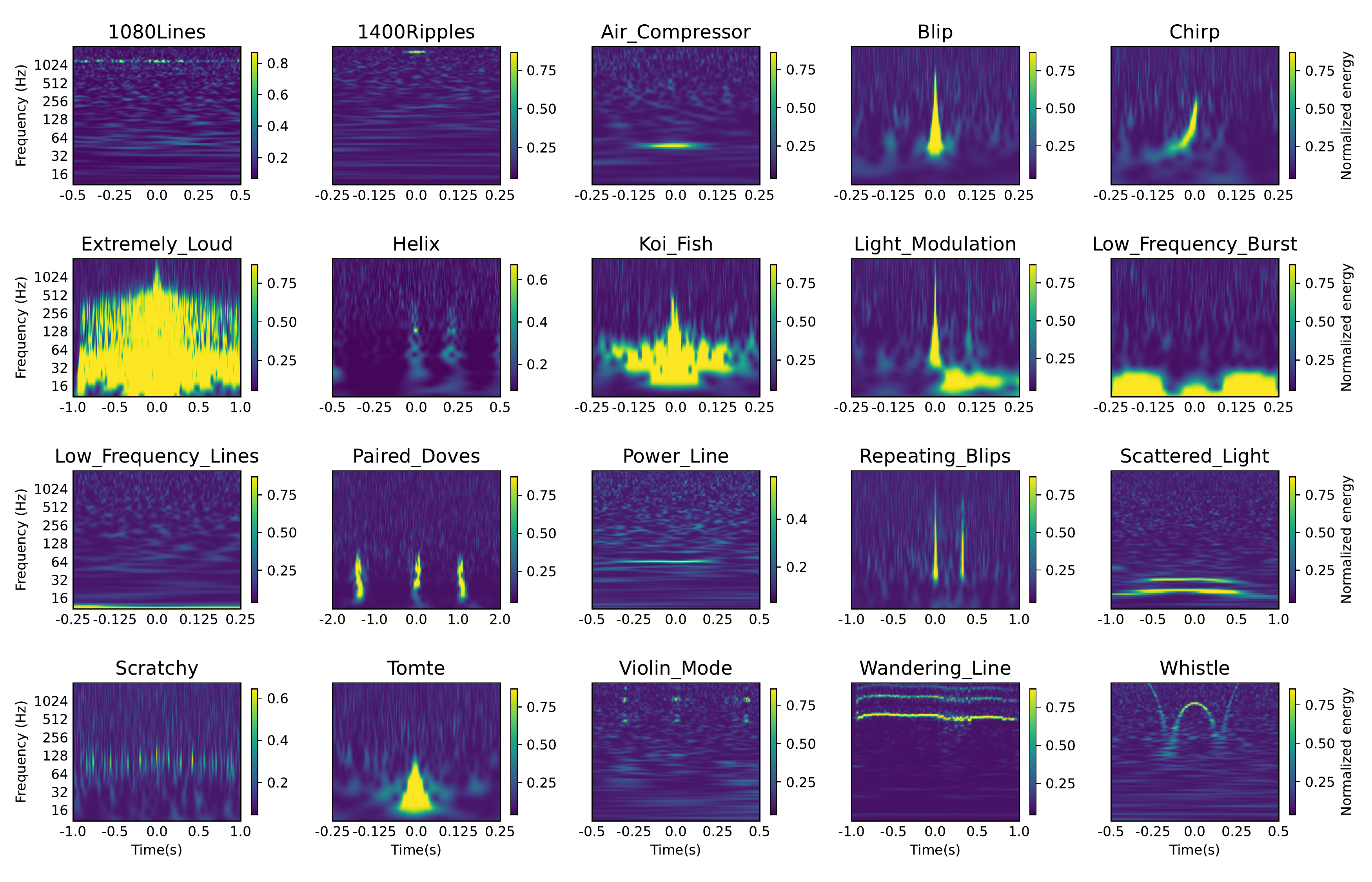}
    \caption{Examples of the different glitch classes from the Gravity Spy dataset. Note that the classes \texttt{None\_of\_the\_Above} and \texttt{No\_Glitch} are not represented in the figure. Moreover, the timescale is not equal for all images and the grey-scale spectrograms are shown with a colour map for better visualization.}
    \label{fig: glitch classes}
\end{figure*}

Following this first training section, transfer learning is applied by removing the last layer of the trained model and replacing it with a randomly initiated layer of the same type. This updated model is then trained for the downstream task, as shown in box (b) of Figure~\ref{fig:ssl}, where the dataset used to feed the neural network already has human-made labels. 
In theory, the pretext task allows the model to learn certain useful patterns from the images, making it so that the downstream model is easier to train.
Usually the pretext task tackles a problem that is a subcategory of the problem associated with the pretext task. This technique is more appropriate if there is a high amount of samples, but only few of those are labelled.


The architecture we chose for the self-supervised approach was the ResNet18, with randomly initialized parameters. This model is trained using a dataset composed of images that only correspond to a specific time-span window. Data augmentation is used to increase the variety in the dataset and avoid overfitting. Similarly to the supervised DL methods discussed before, the AdamW optimizer and cross-entropy loss function are used during training, using the 1cycle training policy and the learning rate finder to select a learning rate.

\subsection{Model evaluation}

The metric we select to compare possible model configurations is the F1 score. This is the harmonic mean of precision (the ratio of correct positive predictions to the total number of positive predictions) and recall (the ratio between the number of correct positive predictions and the total number of positive examples) 
\begin{equation}
    \label{eq: f1 score}
    \textrm{F1 score} = \frac{2} { \frac{1}{\textrm{precision}} + \frac{1}{\textrm{recall}} }  = \frac{\textrm{TP}} {\textrm{TP} + \frac{\textrm{FN} + \textrm{FP}}{2} }.
\end{equation}
Above, TP, FP and FN respectively denote the total number, true positives, false positives and false negatives.

The F1 score can be computed for each class but this raises the issue of having multiple metrics. In order to have one single overall metric, the macro-averaged F1 score is used instead. This metric is calculated by plugging in the macro-averaged precision and macro-averaged recall, which are respectively the precision and recall averaged across all classes, in Eq.~\eqref{eq: f1 score}.

As models that train faster are easier to optimize and can also result in faster inference times, a new metric was created to penalize models that take longer to train:
\begin{equation}
    \label{eq: comb_f1_time}
    \textrm{combined\_F1\_time} = \textrm{F1\_score} - \textrm{total\_runtime} / 3\times 10^4.
\end{equation}
The numerical value in the equation was 
conveniently chosen so that an increase of 
0.2 percentage points in the F1 score will only be accepted if it adds less than 60 seconds to the training time. 
Thus, this metric allows to search for better models without increasing too much the training time.

The metrics were computed on the validation dataset for the candidate models, with only the best configurations being chosen for  evaluation on the test set. The use of three separate subsets in the model training and evaluation process is of utmost importance. The inclusion of an intermediate validation set allows to evaluate different model candidates on this set and select them accordingly based on their performance. Then, the selected models can be evaluated on the test set, which had not been used to guide previous decisions, to provide a more accurate estimation of the models' generalization. Had the intermediate set not been used, the test set would provide an over-confident estimate of the models' performance, as they had been optimized for that same dataset.

\section{Dataset}
\label{section: dataset}

The Gravity Spy dataset, introduced in \cite{Zevin2017a,Bahaadini2018},  
is a collection of spectrograms of (mostly) glitches identified during the O1 and O2 observing runs in the Advanced LIGO detectors. This dataset presents a multi-class classification problem, where the goal is to assign the only correct glitch class to each noise sample. Each sample contains four spectrogram images centered at the same instant in time but with different durations of 0.5, 1.0, 2.0 and 4.0 seconds, all represented as $140 \times 170$ pixels grey-scale images, as seen in Figure~\ref{fig: glitch classes}. In this work we directly use the public images from Gravity Spy. It can be observed that the glitch classes vary widely in duration, frequency range and shape. Nevertheless, some classes can be easily mistaken for each other due to their similar morphology, like the \texttt{Blip} and \texttt{Tomte} classes~\cite{Merritt:2021}, or if they are not visualized using the appropriate time window. For instance, a \texttt{Repeating\_Blips} sample may be identified as a \texttt{Blip} if the window is too small. 

In the version of the dataset used in this work\footnote{https://zenodo.org/record/1476156}, v1.0, there are 8583 glitch samples distributed unevenly over 22 different classes, which are thoroughly described in \cite{Zevin2017a}. The majority of examples are \texttt{Blip}s, with almost 1900 samples, while there are five minority classes (\texttt{Paired\_Doves}, \texttt{Wandering\_Line}, \texttt{Air\_Compressor}, \texttt{Chirp}, and \texttt{None\_of\_the\_Above}) with less than 190 examples (10\% of the number of \texttt{Blip}s), the less represented being the \texttt{Paired\_Doves}, with only 27 examples. 
The \texttt{Chirp} class does not actually represent glitches but instead ``hardware injections'', which are simulated signals created to resemble real GWs from CBCs, and used for the testing and calibration of the detectors~\cite{Bahaadini2018}. 
The Gravity Spy dataset is already split, using stratified sampling to ensure similar distributions in each subset, into training (70\%), validation (15\%) and test (15\%) sets, to facilitate the comparison of different methods. 

\section{Results}
\label{section: results}

\subsection{Supervised models}

We start discussing the results of glitch classification for supervised DL models.
For a quick search for a baseline classifier the ResNet18 and ResNet34 architectures were chosen, with an adapted number of inputs for the first convolutional layer, and using random weight initialization. Both architectures were trained from scratch with \texttt{fastai}'s \texttt{fit\_one\_cycle} routine over 15 epochs with the steepest point from the learning rate finder as the maximum learning rate; with the other hyperparameters set to \texttt{fastai}'s defaults. 

Several approaches were tested regarding the amount and format of the information provided to the model. 
In the simplest approach, the model had only access to one of the four spectrograms for each example, always with the same duration. This corresponds to views labelled single1~(0.5~s), single2~(1.0~s), single3~(2.0 s) and single4~(4.0 s). Thus, these single-view models accepted batches of grey-scale (1-channel) $140 \times 170$ (height $\times$ width) pixels images.
The second approach was a merged view model inspired in \cite{Bahaadini2017}, which consists of placing the four single view $140 \times 170$ images next to each other, forming a $280 \times 340$ pixels image. 
Finally, we implemented an approach based on encoding information related to a different time duration in each image channel, as introduced in \cite{George2017}. Encoded views with all combinations of two to four time durations were compared.

\begin{figure}[t!]
    \centering
    \includegraphics[scale=0.47]{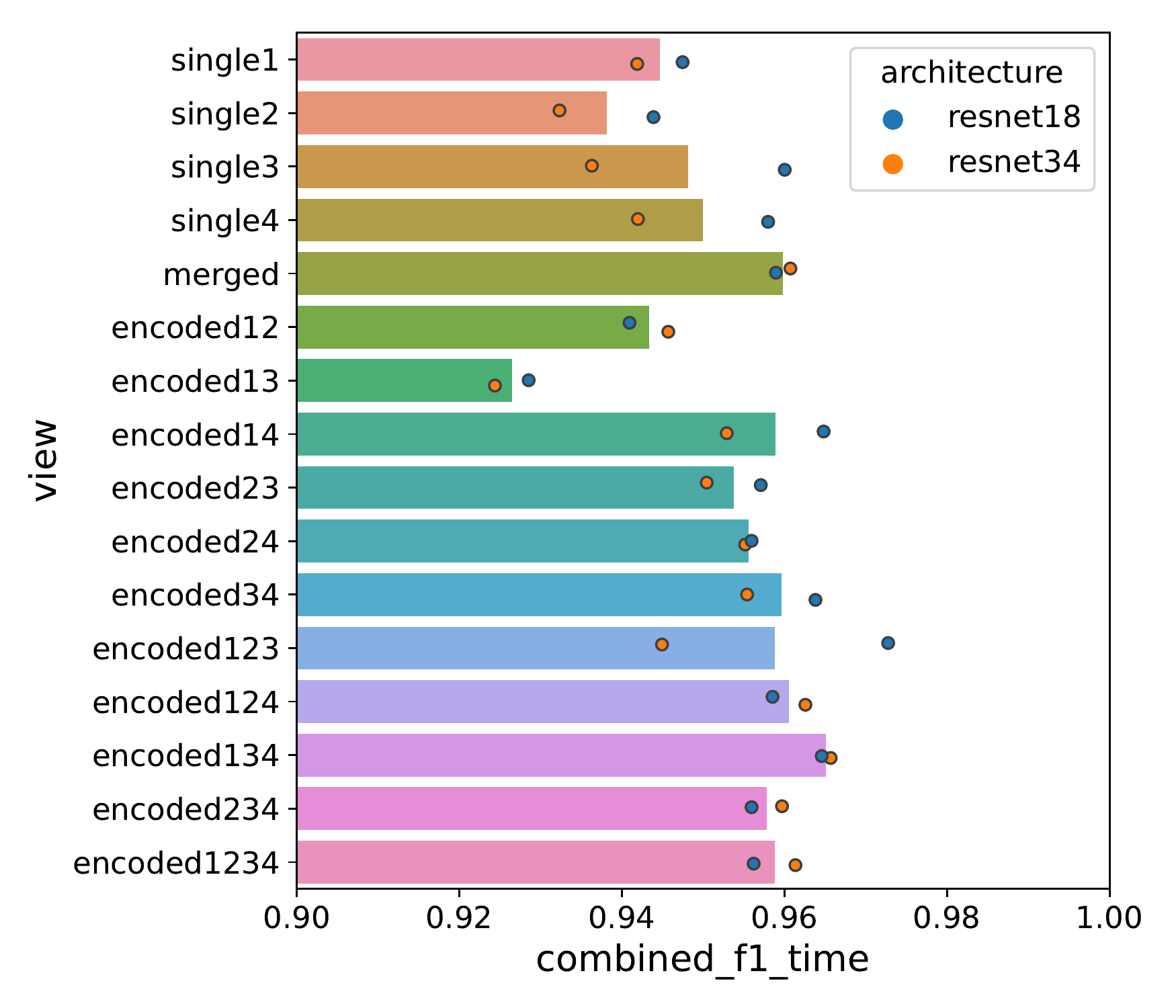}
    \caption{Comparison of the different views regarding combined\_f1\_time computed over the validation set. The horizontal bars show the average of the result of the two architectures, ResNet18 and ResNet34.}
    \label{fig: baseline_scratch_views}
\end{figure}

A hyperparameter grid search was run over the two model architectures and all the possible views, in order to select the best views and choose a baseline model. The results in terms of each view are shown in Figure~\ref{fig: baseline_scratch_views}.
It was found that the single views were outperformed by almost all the other views.
The highest average F1 scores were obtained by encoded134. Remarkably, this outperformed merged view and encoded1234, which use all the available information. 
Taking the total training time in consideration using the combined\_f1\_time metric from Eq.~\eqref{eq: comb_f1_time}, we conclude that encoded134 is the best view, as the merged view takes about 6 more seconds per epoch (a 70\% increase) than view encoded134, while achieving slightly worse performance regarding the F1 score. Therefore, encoded134 was selected as the view used for the rest of the supervised models.

\begin{figure}[!t]
    \centering
    \includegraphics[scale=0.55] {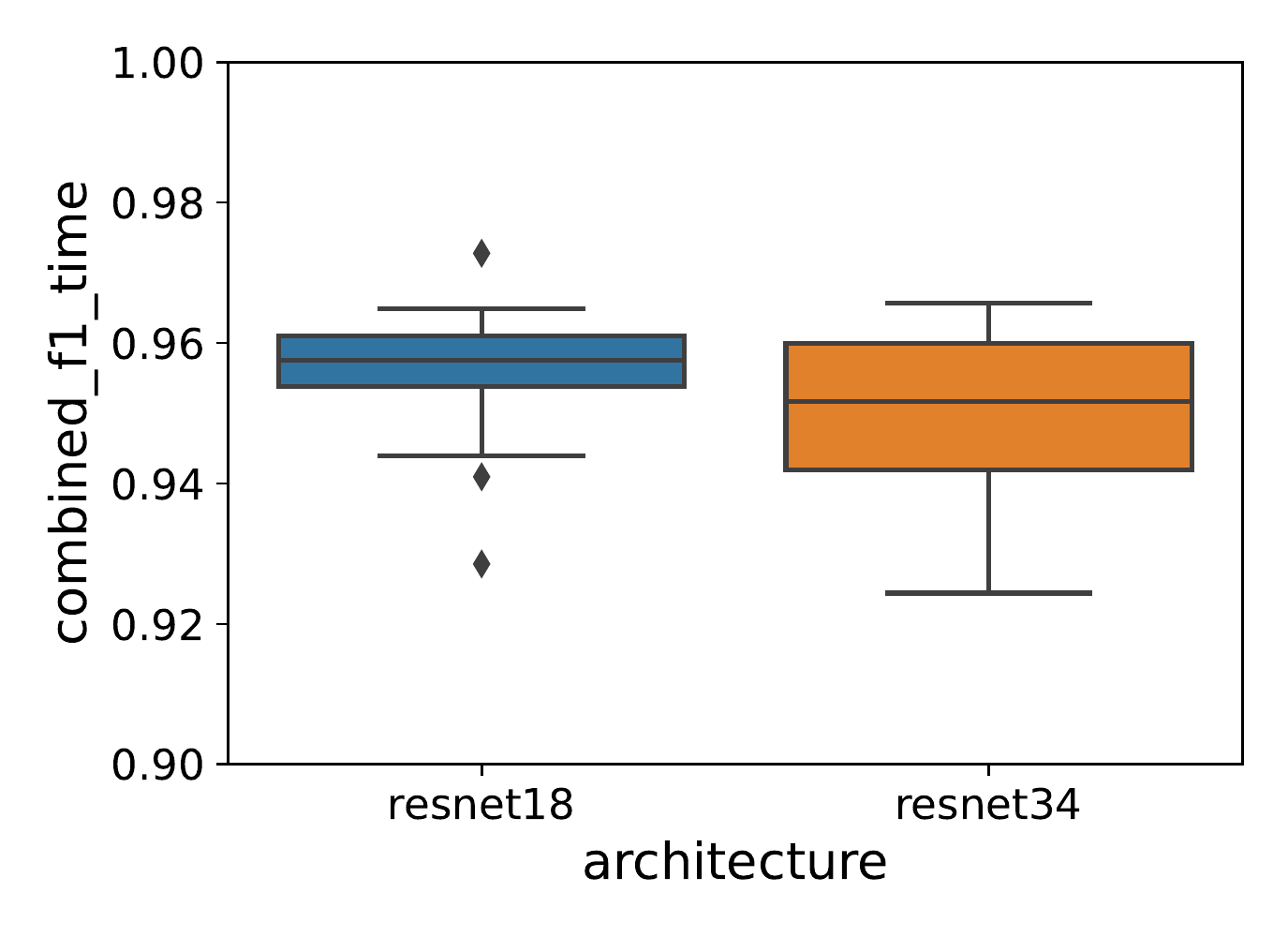}
    \caption{Box plots with the comparison of the architectures regarding combined\_f1\_time, computed over the validation set.}
    \label{fig: baseline_scratch_architectures}
\end{figure}

In addition, the results of the grid search were also evaluated in terms of the chosen architecture, as depicted in 
Figure~\ref{fig: baseline_scratch_architectures}. This figure shows box plots of the $\textrm{combined\_F1\_time}$ for the validation set\footnote{In box plots (which we employ in several figures in this paper) the lower limit of the box corresponds to the $25^\mathrm{th}$ percentile of the data, and the upper limit to the $75^\mathrm{th}$ percentile. The line which goes through the box represents the median of the observations. The whiskers represent the maximum and minimum (excluding outliers) of the quantity being displayed, and the individual points (if any) are outliers, i.e.~points that are more than 1.5 inter-quartile distance away from the closest quartile.}. We found that, on average, ResNet18 achieves better results both on the F1 score and the training time, resulting in a median combined\_F1\_time 0.6 percentage points higher than ResNet34. Thus, the ResNet18 was chosen for the baseline model.

After deciding the baseline configuration, five models were trained for 15 epochs each, using the same configuration as before, and selecting the ResNet18 architecture and the encoded134 view. The F1 scores of the five training runs range between 96.7\% and 98.1\%, and the best one was chosen as the baseline model. 

An attempt was made to tune the learning-rate choice method, the number of epochs and the batch size, with the goal of further optimizing the model's performance. Hence, a Bayesian sweep was performed over 50 configurations of the following three hyperparameters: (a) number of epochs, between 8 and 25; (b) batch size, sampling from the set $[32, 64, 128, 256]$; and (c) learning rate finder function, sampling from the four available methods in \texttt{fastai} (\texttt{steep}, \texttt{valley}, \texttt{slide}, \texttt{minimum}).
It was found that no model obtained an F1 score higher than the baseline, although some models reached a slightly better combined\_f1\_time score by maintaining a comparable F1 score while training for less epochs.

\begin{figure}[!t]
    \centering
    \includegraphics[scale=0.48]{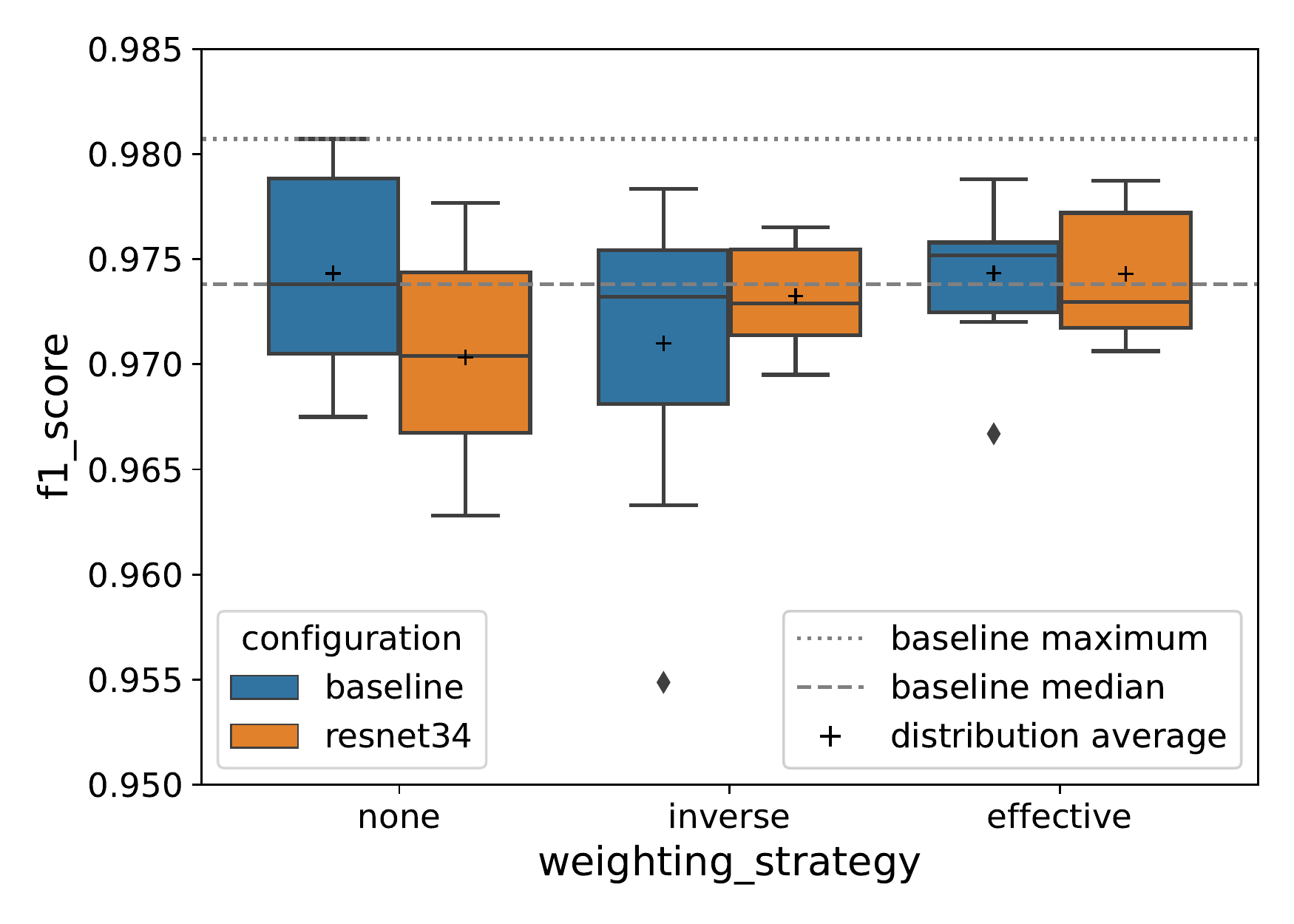}
    \caption{Box plots of the F1 scores on the validation set for three different weighting strategies using three different configurations.}
    \label{fig: compare weighted}
\end{figure}

The weighted loss function from Eq.~\eqref{eq: cross entropy reweighted} was also tried, in order to further penalize the model for mistakes in the least represented classes. The re-weighting strategies were compared using the baseline configuration in addition to a configuration equal to the baseline but with the ResNet34 architecture instead of the ResNet18. Each configuration/strategy combination was run for 10 times, yielding the  results shown in Figure~\ref{fig: compare weighted}. 
When the inverse weighting strategy is introduced, the performance of the baseline decreases, while the ResNet34's F1 scores slightly increase. 
The effective re-weighting strategy seems to result in similar performance for the baseline in comparison to the models without re-weighting, while the ResNet34 models, once again, slightly improve their performance.

\begin{figure}[!t]
    \centering
    \includegraphics[scale=0.46, clip, trim={0cm 0cm 0cm 0cm}]{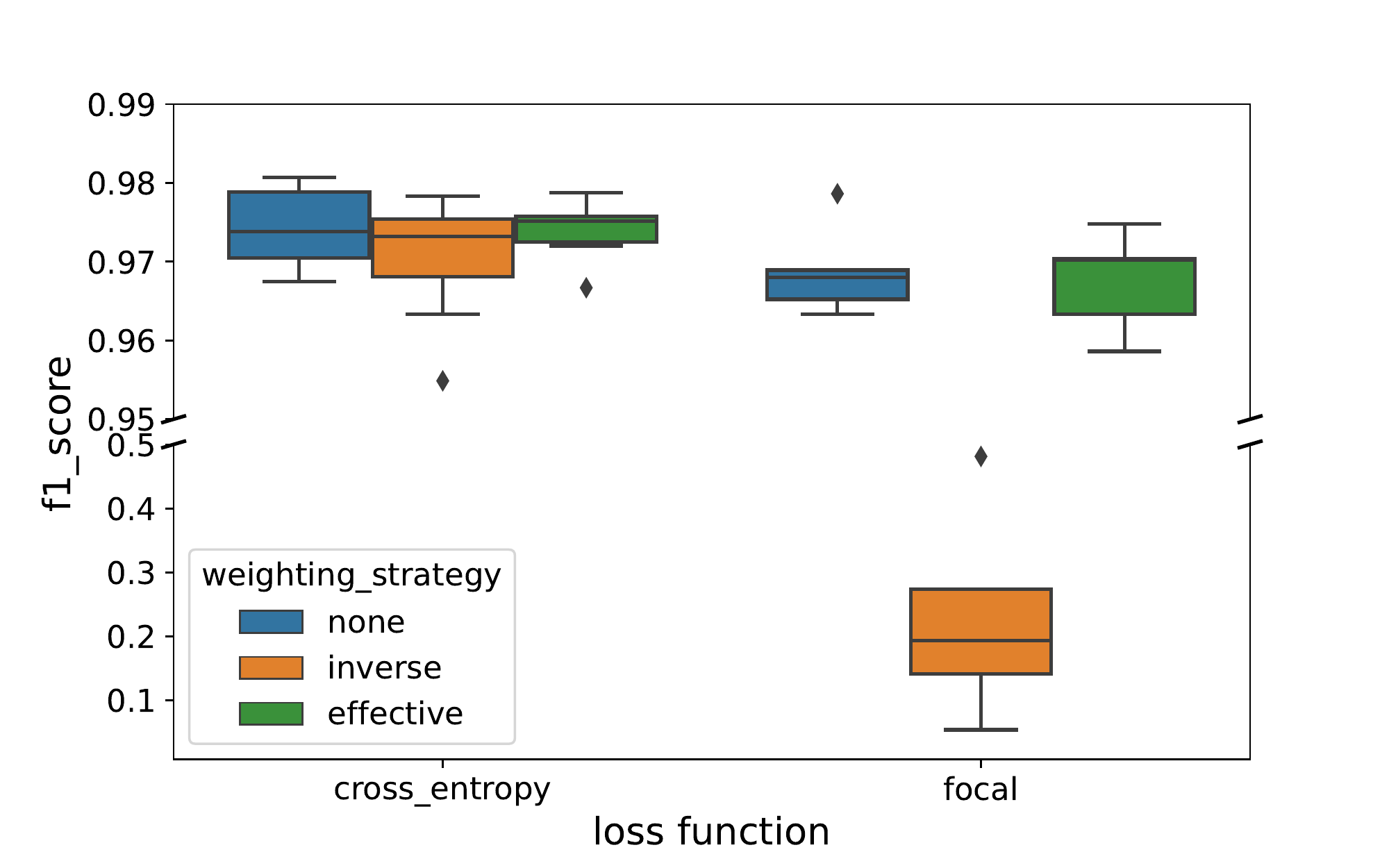}
    \caption{Box plots of the F1 scores on the validation set for the cross-entropy loss and focal loss functions, using the baseline configuration with three different re-weighting strategies.}
    \label{fig: compare focal}
\end{figure}

Next, the focal loss from Eq.~\eqref{eq: focal loss} was tried by training the baseline configuration using this loss five times for each of the previous weighting strategies (including the absence of re-weighting). The results are shown in Figure~\ref{fig: compare focal}. Focal loss does not seem to work with the inverse re-weighting strategy and yields similar performance without re-weighting and using the effective number of samples strategy. The results with focal loss are worse than the results with cross-entropy in all cases.

\subsection{Supervised models with transfer learning}

\begin{figure}[!t]
    \centering
    \includegraphics[scale=0.43]{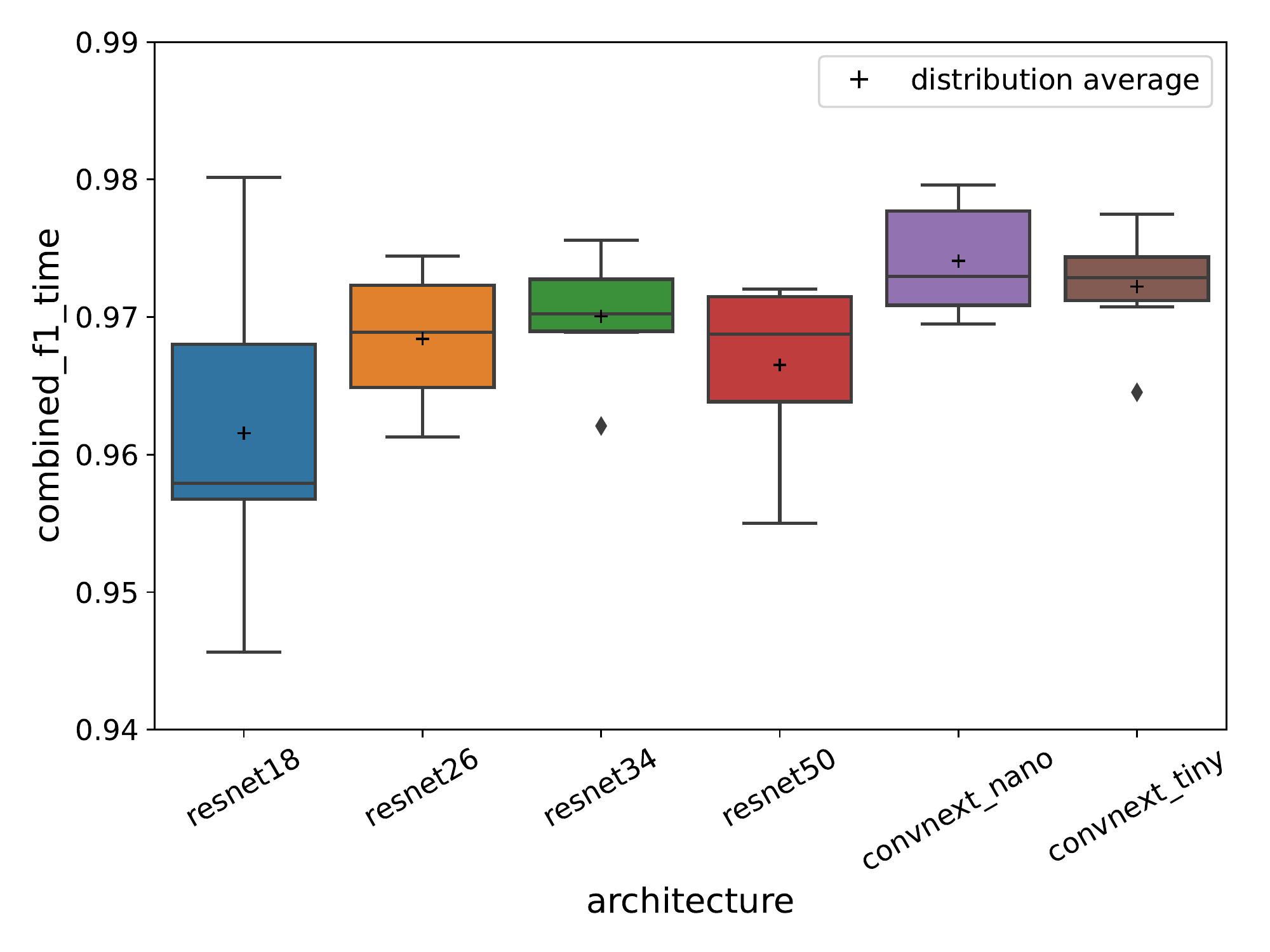}
    \caption{Box plots with the comparison of the architectures regarding combined\_f1\_time, computed over the validation set.}
    \label{fig: transfer models arch}
\end{figure}

After being unable to improve the baseline model while training from scratch, an approach using transfer learning was attempted.
Initially, models pre-trained on the ImageNet dataset using the ResNet18 and ResNet34 architectures were trained with \texttt{fastai}'s \texttt{fine\_tune} routine with the same configuration as the baseline model. As the results suggested that the ResNet18 architecture may not be the best one for transfer learning, other architectures from the ResNet family were also tried (ResNet26 and ResNet50) as well as architectures from the ConvNeXt family (ConvNeXt\_Nano and ConvNeXt\_Tiny).
Having chosen these six architectures, a grid sweep was performed in order to compare them. In particular, we trained each model for 4, 9 and 14 epochs (in addition to the initial frozen epoch), using two different learning rate functions, \texttt{steep} and \texttt{minimum}, while keeping the other hyperparameters at the default values. In total, each architecture was trained six times, with the results shown in Figure~\ref{fig: transfer models arch}. 
It was observed that ResNet18 produced the best single run, but it appears to be worse than all the other architectures when averages and medians over multiple training runs are considered. The performance of the ResNet family seems to peak with the ResNet34 architecture but it is surpassed, on average, by the ConvNeXt family. ConvNeXt\_Nano is close to ConvNeXt\_Tiny in terms of F1 score but is preferable when the training time is considered, resulting in higher combined\_f1\_time values. For these reasons, ConvNeXt\_Nano was chosen as the architecture for subsequent optimizations. 

A Bayesian sweep was then performed to optimize the hyperparameters of the transfer learning model, with the ConvNeXt\_Nano architecture. Fifty configurations were evaluated, given the following parameter space: (a) number of frozen\_epochs, with a uniform integer distribution between 1 and 5; (b) number of unfrozen epochs, using a uniform integer distribution between 1 and 10; (c) batch size, sampling from the set $[32, 64, 128, 256]$; (d) learning rate finder function, sampling from the four available methods in \texttt{fastai} (\texttt{slide}, \texttt{valley}, \texttt{steep}. \texttt{minimum}); (e) loss function: either cross-entropy or focal loss; and (f) re-weighting strategy, which could be none, inverse of class frequency, or ENS.  

\begin{figure}[!t]
    \centering
    \includegraphics[scale=0.47]{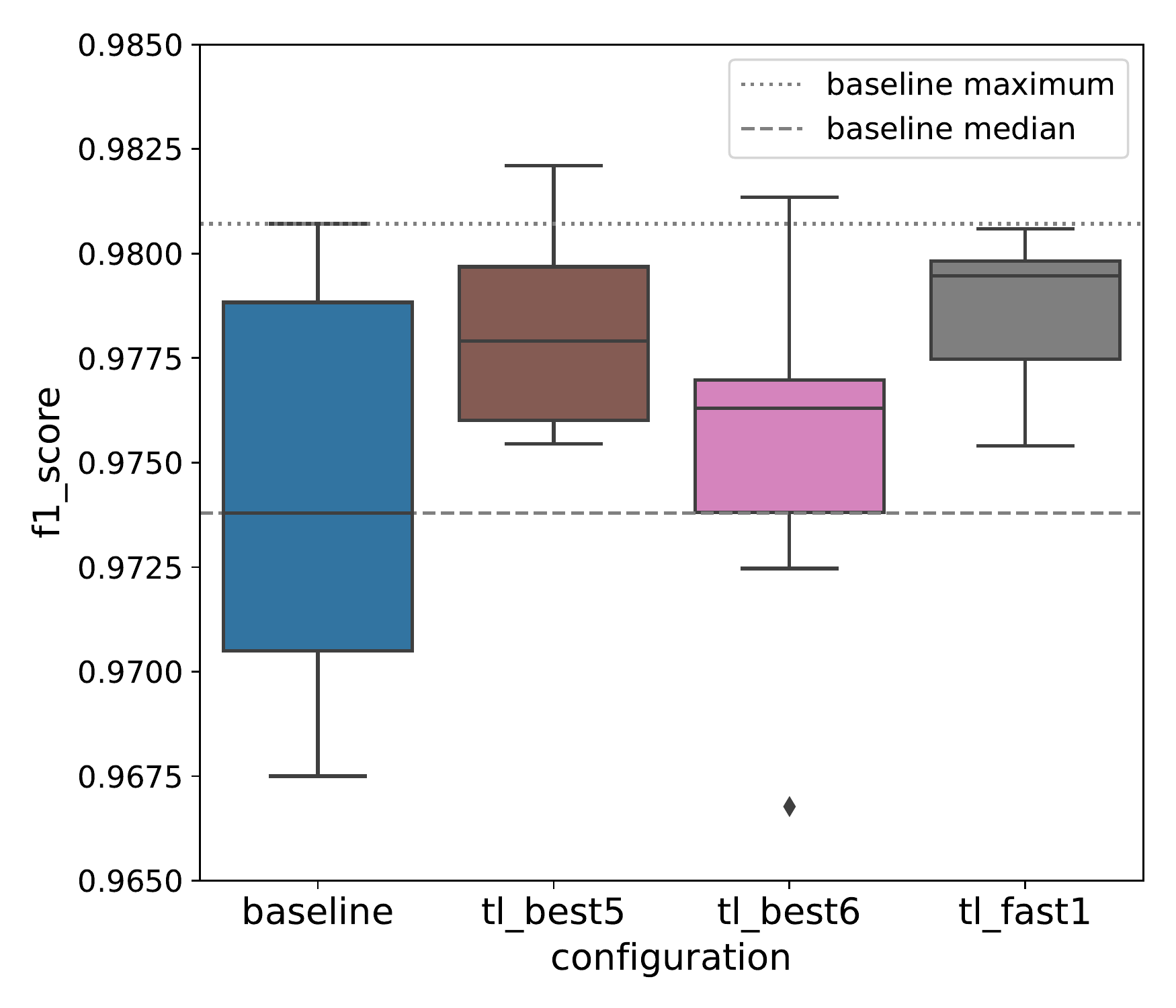}
    \caption{Box plots of the F1 scores on the validation set of the ten runs for the baseline and the best three configurations selected from the second Bayesian sweep.}
    \label{fig: transfer optimized scores}
\end{figure}

The six overall best configurations of the sweep as well as the three best models trained for 6 epochs or less were more thoroughly evaluated by re-training them five times each. 
From these, the three most promising configurations in terms of F1 score were selected, which have the following hyperparameters: (a) \texttt{tl\_best5}: 2 frozen and 8 unfrozen epochs, batch size of 64, \texttt{minimum} learning rate selection function and cross-entropy loss with inverse class re-weighting; (b) \texttt{tl\_fast1}: the same hyperparameters as \texttt{tl\_best5} except for a quicker training with 1 frozen and 5 unfrozen epochs; and (c) \texttt{tl\_best6}: 1 frozen and 9 unfrozen epochs, batch size of 256, \texttt{steep} learning rate selection function and focal loss with ENS re-weighting. They were trained five more times, totaling ten models for each, and the results are shown in Figure~\ref{fig: transfer optimized scores}.

The highest score was obtained by the best run of the \texttt{tl\_best5} configuration, which beat the baseline's best run. 
The \texttt{tl\_best5} and \texttt{tl\_fast1} configurations achieved F1 scores higher than the baseline median in all runs. Moreover, \texttt{tl\_fast1} has the highest median among all models.

The model with the highest overall score, corresponding to the best run of \texttt{tl\_best5}, was chosen as the transfer learning classifier. In comparison with the baseline model, it only increases the F1 score from 98.07\% to 98.21\%, but the transfer learning configuration appears to result in more stable performances across multiple training trials, with its worst performance higher than the baseline median.
In fact, the baseline configuration achieves an F1 score of $(97.4 \pm 0.5)\%$, less than \texttt{tl\_best5}'s $(97.8 \pm 0.2)\%$ and \texttt{tl\_fast1}'s $(97.9 \pm 0.2)\%$. Note that the previous F1 scores are reported in the form $(\mu \pm \sigma) \%$, where $\mu$ is the distribution average and $\sigma$ its standard deviation. 
We highlight that these models were very quick to train, with training times between 1 and 2 minutes using a NVIDIA RTX A4000 GPU (16 GB VRAM). 

\subsection{Self-supervised model}

In contrast to supervised learning, self-supervised learning is a two-phase process that partly uses automatically generated labels, i.e.~pseudo-labels, during the training of the architecture. This can be considered an advantage for datasets where data is abundant but the labeling of the samples is scarce or incorrect. As stated in Section \ref{subsection: methods self-sup}, the self-supervised learning framework first needs to solve the pretext task and then the downstream task. This means that this approach utilizes two distinct training phases, each one using its own independent dataset. The first training phase gives rise to the first model (I), which solves the pretext task. In the second phase, we apply transfer learning to the previous model, to obtain the second and final model (II), which solves the downstream task.

After applying the image transformations (in the first phase) or image augmentations (in the second phase), all the images for both datasets are then resized to 128$\times$128 pixels through interpolation. This is done to make the models train faster since the input's size is smaller. Just like in supervised models, the F1 score and accuracy are the metrics being examined. Both metrics are only applied to the validation dataset. In both phases the batch size is set to 64.

Model (I) was trained from scratch using pseudo-labels, with a dataset composed of single view images from  all four time-span windows (0.5 s, 1.0 s, 2.0 s, and 4.0 s). Each signal has thus four independent single view images associated with it. The visual transformation is picked from a list of operations like random crops (\texttt{crop}), horizontal and vertical flips (\texttt{flipVertical} and \texttt{flipHorizontal}), color jitters (\texttt{colorize}), and rotations of 90º and 270º (\texttt{rotate90} and \texttt{rotate270}). The no-transformation case was also used (labeled as \texttt{noTransformation}). For each image, one visual transformation is randomly selected from the list. Once the visual transformation is applied, a pseudo-label is associated with it. For instance, if the vertical flip option is selected from the list, the resulting image will be flipped vertically and then resized to 128 pixels by 128 pixels. The resulting pseudo-label is denominated as the \texttt{flipVertical} class.

For simplicity, model (II) only utilizes data with 4-second time windows,  although we note that certain types of glitches require longer windows to be recognized. Similar to the previous dataset, all images are single views. As a consequence, the dataset used to train model (I) has four times more images for each epoch compared to the dataset from model (II). During the training phase of this model, the original labels are used instead of the pseudo-labels. In this phase, data augmentation is applied to avoid overfitting, as explained below. Before resizing the image, each one is rotated between -5 and 5 degrees, then undergoes a color jitter, and finally a random crop is applied. Note that since the original labels are used, there is no longer a correlation between the class and the type of augmentation applied to the image.

For both models, the learning rate finder is used to determine the values of the learning rate. For model (I) the learning rate is $6.92\times10^{-3}$, and for model (II) is $5.75\times 10^{-4}$. Since the 1cycle policy is used during training, these values actually correspond  to the maximum learning rate, instead of a single fixed learning rate value for all epochs.

\begin{figure}[!t]
    \centering
    \includegraphics[scale=0.28]{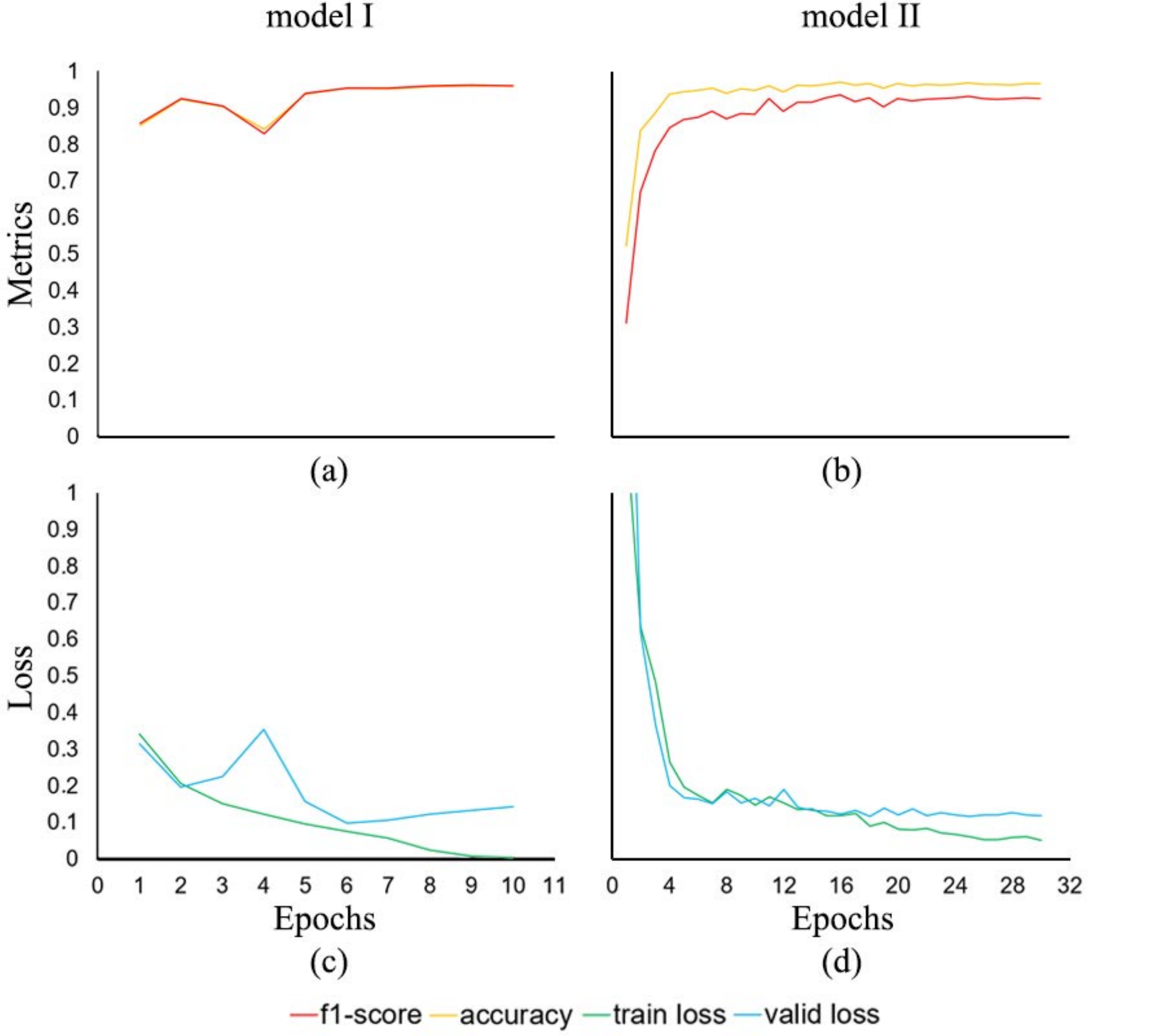}
    \caption{Epoch evolution of the accuracy and macro-averaged F1 score for training of model I (a) and model II (b). The corresponding losses are represented in panels (c) and (d).}
    \label{fig: ssl_metrics_loss}
\end{figure}

Model (I) was trained during 10 epochs since both the values of the loss and the metrics quickly converged after the first few epochs, as shown in Figure \ref{fig: ssl_metrics_loss}(a) and (c). 
Model (II), however, was trained for 30 epochs. The evolution of the metrics for this model, as well as the loss, are represented in Figure~\ref{fig: ssl_metrics_loss}(b) and (d). The values start to converge somewhere between the $10^\mathrm{th}$ and the $15^\mathrm{th}$ epoch. In the first five epochs, only the parameters of the last layer were allowed to change with a maximum learning rate determined by the learning rate finder. In the remaining 25 epochs, all architecture parameters can be modified, with the maximum learning rate being $5.75\times 10^{-5}$, a 10 times smaller rate than in the first five epochs. The resulting final model (II) has an F1 score of 92.74\% and an accuracy of 96.74\%, evaluated on the validation dataset.

\subsection{Evaluation of the best models on the test set}

The best models found in the previous sections are now evaluated on the test dataset, for both supervised and self-supervised approaches. It is worth recalling that this dataset was not used during the training of the neural networks.

\begin{figure*}[!htp]
    \centering
    \includegraphics[scale=0.63]{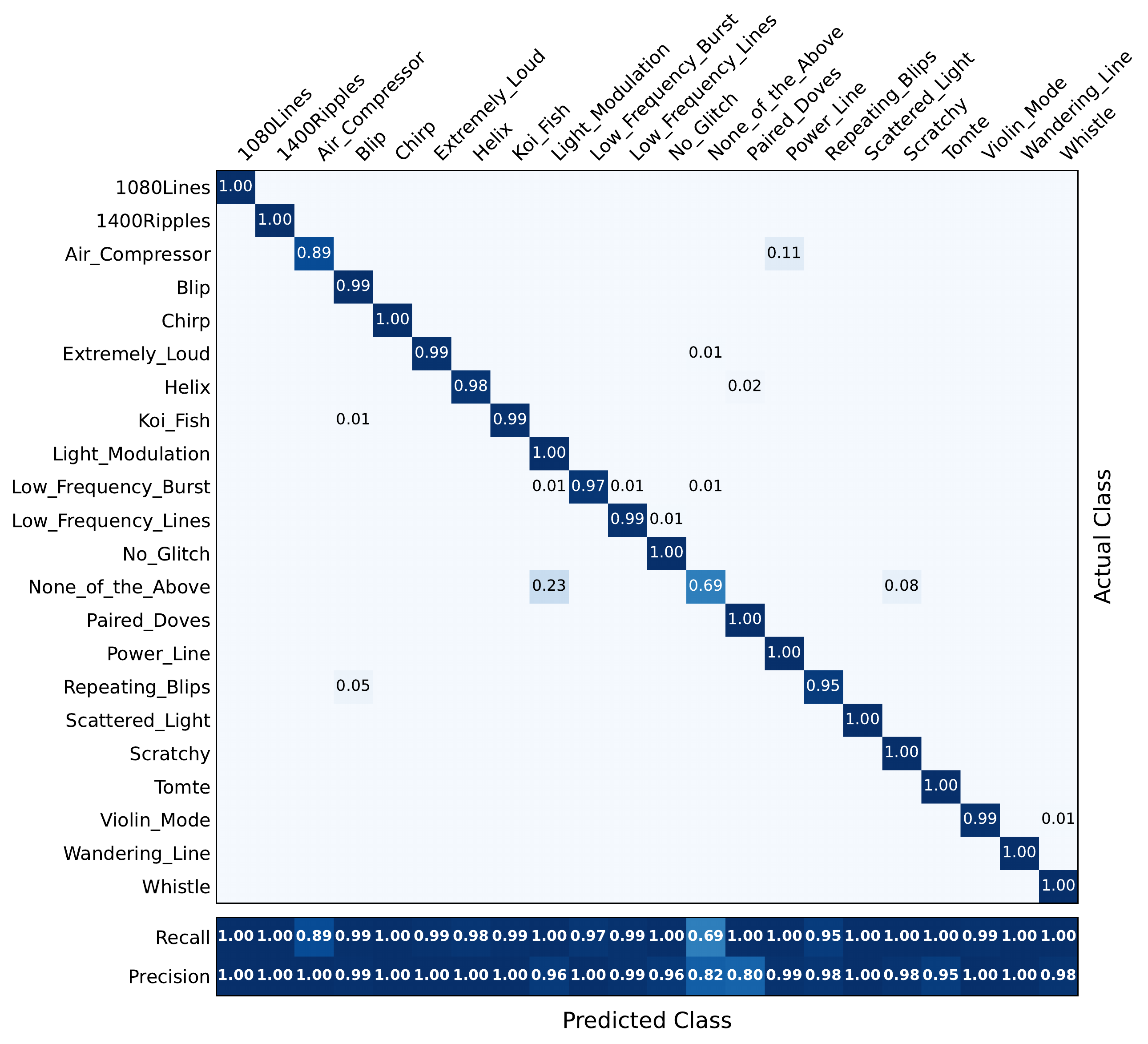}
    \caption{Normalized confusion matrix of the predictions of the baseline model over the test dataset, for the supervised approach.}
    \label{fig: baseline_test_cm}
\end{figure*}

From the supervised models, both the baseline and \texttt{tl\_best5} models were selected to be evaluated on the test dataset, based on their performance on the validation dataset. 
We found that the F1 score of the baseline slightly dropped to 97.18\%, while \texttt{tl\_best5}, the best model on the validation dataset, performed worse than the baseline on the test dataset, with an F1 score of 96.84\%. 
Nevertheless, the F1 scores obtained on data which was never seen by the models nor used to perform choices regarding the hyperparameter optimization, are still very high for both models. The confusion matrix of the baseline model, using the predictions on the test dataset, is presented in Figure~\ref{fig: baseline_test_cm}. 
Both precision and recall are, at least, 95\% for 19 out of the 22 classes of glitches. The model appears to have more difficulties in the \texttt{Air\_Compressor} and \texttt{None\_of\_the\_Above} classes. This may be explained by the fact that, for these classes, 
the number of images in the training set is significantly smaller than for the other classes.
Note that the \texttt{None\_of\_the\_Above} class, in particular, has a very diverse distribution since it encompasses every glitch outside the defined classes. This class has been deprecated from the glitch classes for O3 data~\cite{Glanzer:2023}.

\begin{figure*}[!htp]
    \centering
    \includegraphics[scale=0.63]{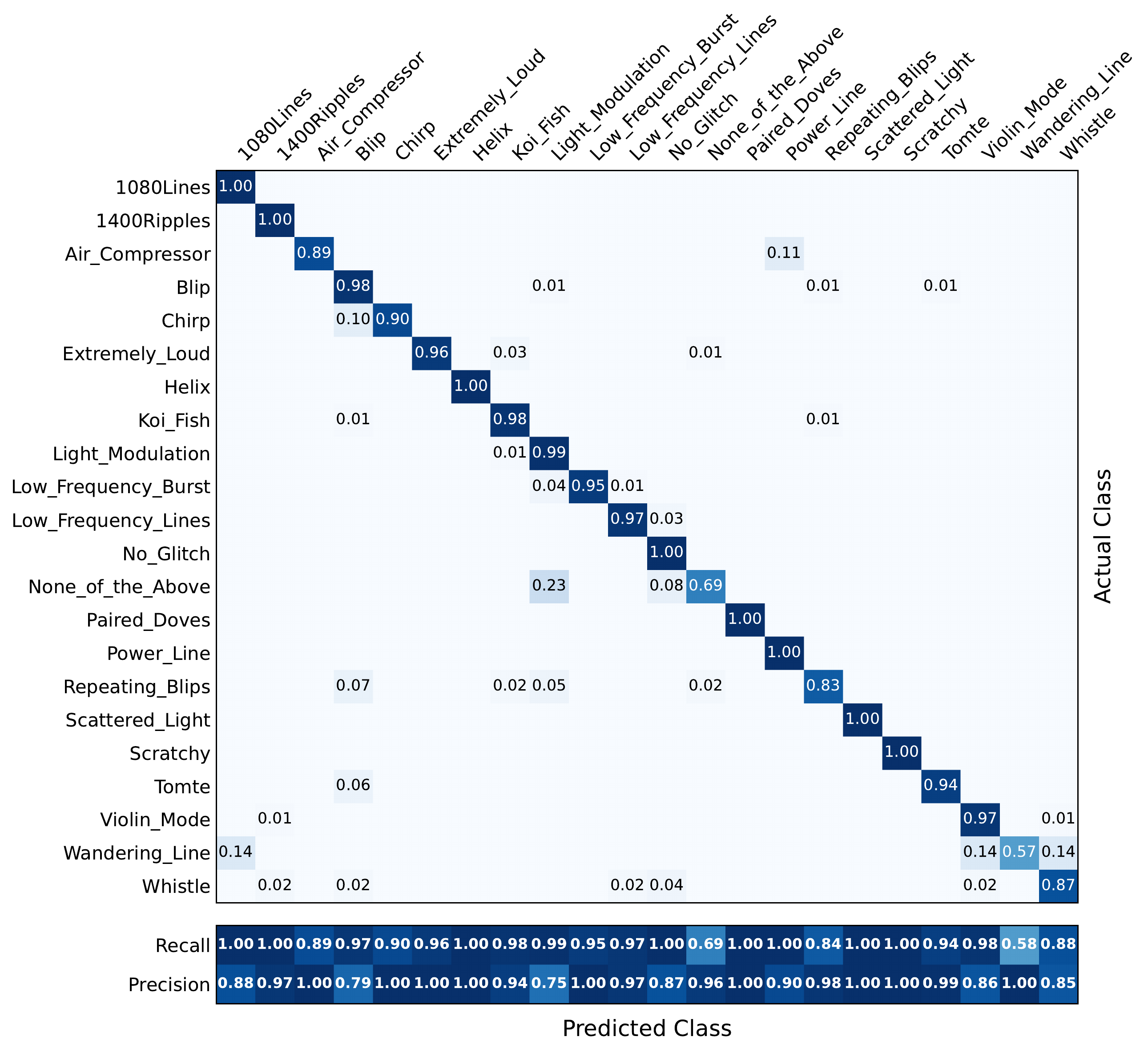}
    \caption{Normalized confusion matrix of the predictions of the self-supervised trained model (II), for the test dataset.}
    \label{fig: ssl_cm_2}
\end{figure*}

\begin{figure}[!htp]
    \centering
    \includegraphics[scale=0.65]{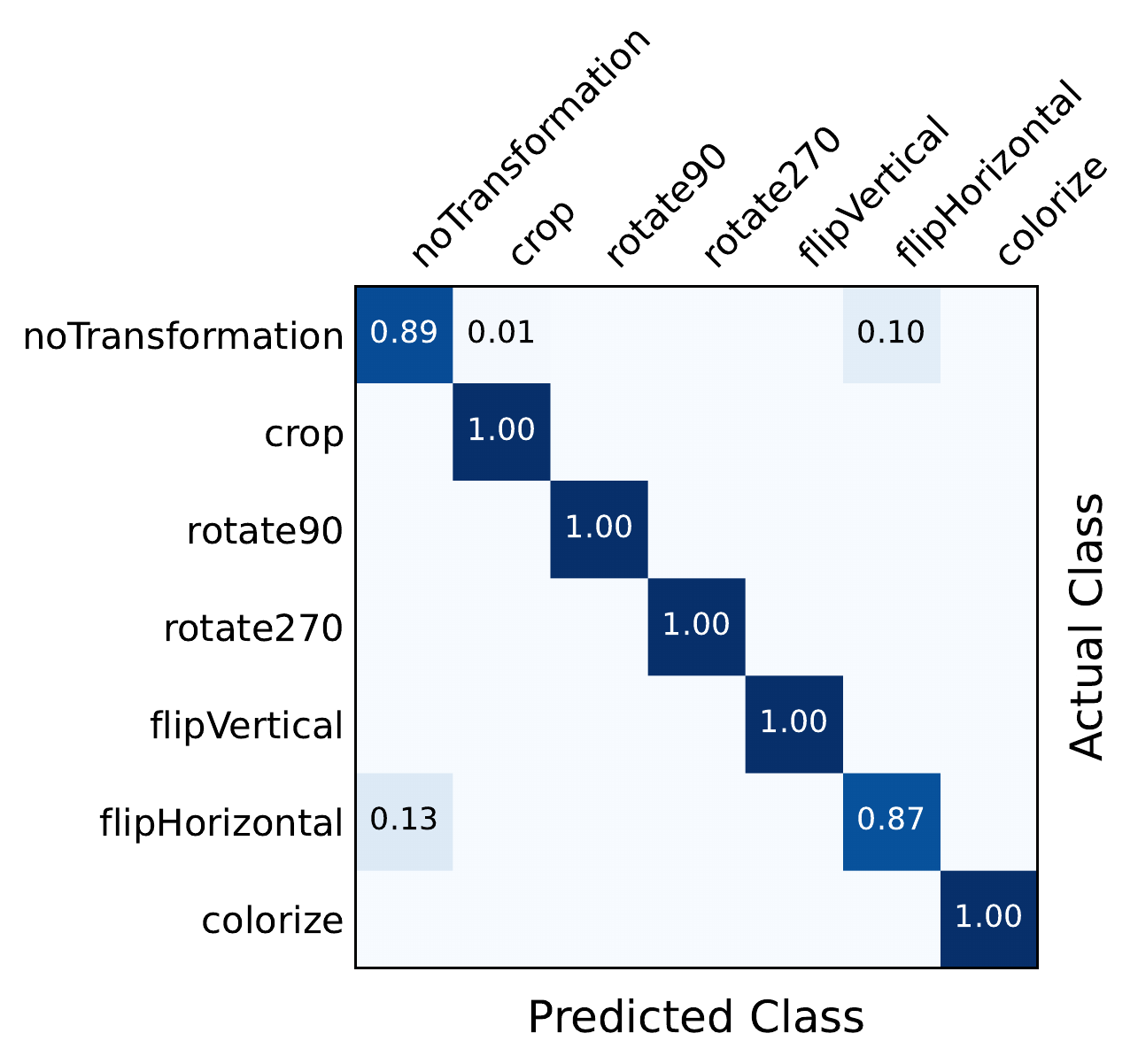}
    \caption{Normalized confusion matrix of the predictions of the self-supervised trained model (I) over the test dataset, before transfer learning is applied}
    \label{fig: ssl_cm_1}
\end{figure}

For the self-supervised model, the corresponding test dataset confusion matrix is shown in Figure~\ref{fig: ssl_cm_2}. The model shows that the recall and precision scores are, at least, 95\% for 14 out of the 22 glitch classes. This is slightly worse than the result obtained under the supervised approach, as expected. In particular, the model struggles to identify several other classes beyond those already signaled for the supervised approach above, like the \texttt{Wandering\_Line} and \texttt{Repeating\_Blips} classes.
This happens due to the fact that, for simplicity, these models were only trained with datasets consisting of 4-second wide signals in contrast to some of the previous models that were trained with encoded views.
As a result of the resized augmentation, the performance of the model might also be affected, as reducing the number of pixels will reduce the image's details. For these reasons, this model did not perform as well as the supervised trained models. However, the performance is still high, with an overall accuracy of 96.50\% (relative to all classes) and an macro-averaged F1 score of 94.15\% on the test dataset. 
For completeness, the confusion matrix of the model trained with pseudo-labels, before transfer learning is applied (first training phase), is also shown in Figure~\ref{fig: ssl_cm_1}. We note that the model has some difficulties distinguishing between the pseudo-labels \texttt{resize} and \texttt{flipHorizontal}, which makes sense since some glitches have vertical symmetry.

The two best supervised models, as well as the best self-supervised model, were compared to the literature in terms of their F1 score and accuracy. The results are shown in Table~\ref{table: compare literature}. 
Both supervised models obtained competitive results, with accuracies higher than the merged view CNNs reported in~\cite{Bahaadini2017,Bahaadini2018}, and the baseline model's performance was also better than the hard fusion ensemble of CNNs~\cite{Bahaadini2018}. 
This model, which uses a ResNet18 architecture trained from scratch, is only slightly worse than the fine-tuned ResNet50 from~\cite{George2018}, with a decrease in F1 score of less than 0.2 percentage points. It should be stressed, however, that Ref.~\cite{George2018} used a different dataset split, without a validation dataset, resulting in more training data and a different test set. More importantly, the absence of a validation set could imply that the results reported by~\cite{George2018} are over-confident. They report the test set performance from the epochs selected because they had the higher test set performance themselves. 
On the other hand, our results from self-supervised model scores show slightly lower values with respect to supervised approaches but the reported numbers are still interestingly high and promising, as they are close to the best performance achieved in Ref.~\cite{Bahaadini2018} using a single view.

\begin{table*}[htp!]
    \renewcommand{\arraystretch}{1.2}
    \caption{Performance of different models on the Gravity Spy dataset.}
    \label{table: compare literature}
    \centering
    \begin{tabular}{ l @{\qquad} r @{\qquad} r @{\qquad} l}
        \hline
        Model & F1 score (\%) & accuracy (\%) & Notes \\
        \hline
                single2 view CNN \cite{Bahaadini2018} & not reported & 96.81      &  custom shallow CNN using only one view \\
               merged view CNN \cite{Bahaadini2017} & not reported & 96.89      &   different dataset version (20 classes)  \\
               merged view CNN \cite{Bahaadini2018} & not reported &    97.67   &  improved version of \cite{Bahaadini2017} \\
               hard fusion ensemble \cite{Bahaadini2018} & not reported &  98.21     &  combines four CNNs  \\
               fine-tuned ResNet50 \cite{George2017, George2018} & 97.65 &  98.84  &    different split (no validation set)  \\
               baseline [this work] & 97.18 & 98.68      & ResNet18 trained from scratch  \\
               tl\_best5 [this work] & 96.84 & 98.14      & fine-tuned ConvNeXt\_Nano  \\
               self-supervised model [this work] & 94.15 & 96.50 & self-supervised trained ResNet18\\
        \hline
    \end{tabular}
\end{table*}

\subsection{Evaluation of the best models on GW signals}

As mentioned before, the Gravity Spy dataset contains a few examples of the \texttt{Chirp} class, which are simulated GW signals from CBCs. In addition, the best models have shown perfect precision and recall for chirp classification in the Gravity Spy test dataset (cf.~Figure \ref{fig: baseline_test_cm}).
These two reasons motivated us to conduct a brief investigation of whether the models trained in this work would be able to identify real detections of GWs as \texttt{Chirps}, using data from the Advanced LIGO detectors collected during the third LVK observing run~(O3).

The strain data for the confident detections in the O3 run, for each Advanced LIGO detector, Hanford (``H1") and Livingston (``L1"), was obtained from the Gravitational Wave Open Science Center (GWOSC)~\cite{Abbott2021}.
For each data sample, a Q-graph of the strain data in the 5 seconds around the GPS time of the detection was obtained using the \texttt{q\_transform} method from the \texttt{GWpy} library~\cite{gwpy}. Furthermore, four crops of this Q-graph, centered at the time with the highest energy, were obtained, one for each different single view duration in the Gravity Spy dataset.
From the resulting 178 samples (89 events times, 2 detectors), many were discarded due to the GW signal being buried in the background noise, with no visible chirp, resulting in 50 samples with clearly visible chirps.

\begin{figure*}[!htp]
    \begin{subfigure}[b]{0.495\textwidth}
        \centering
        \includegraphics[width=\linewidth]{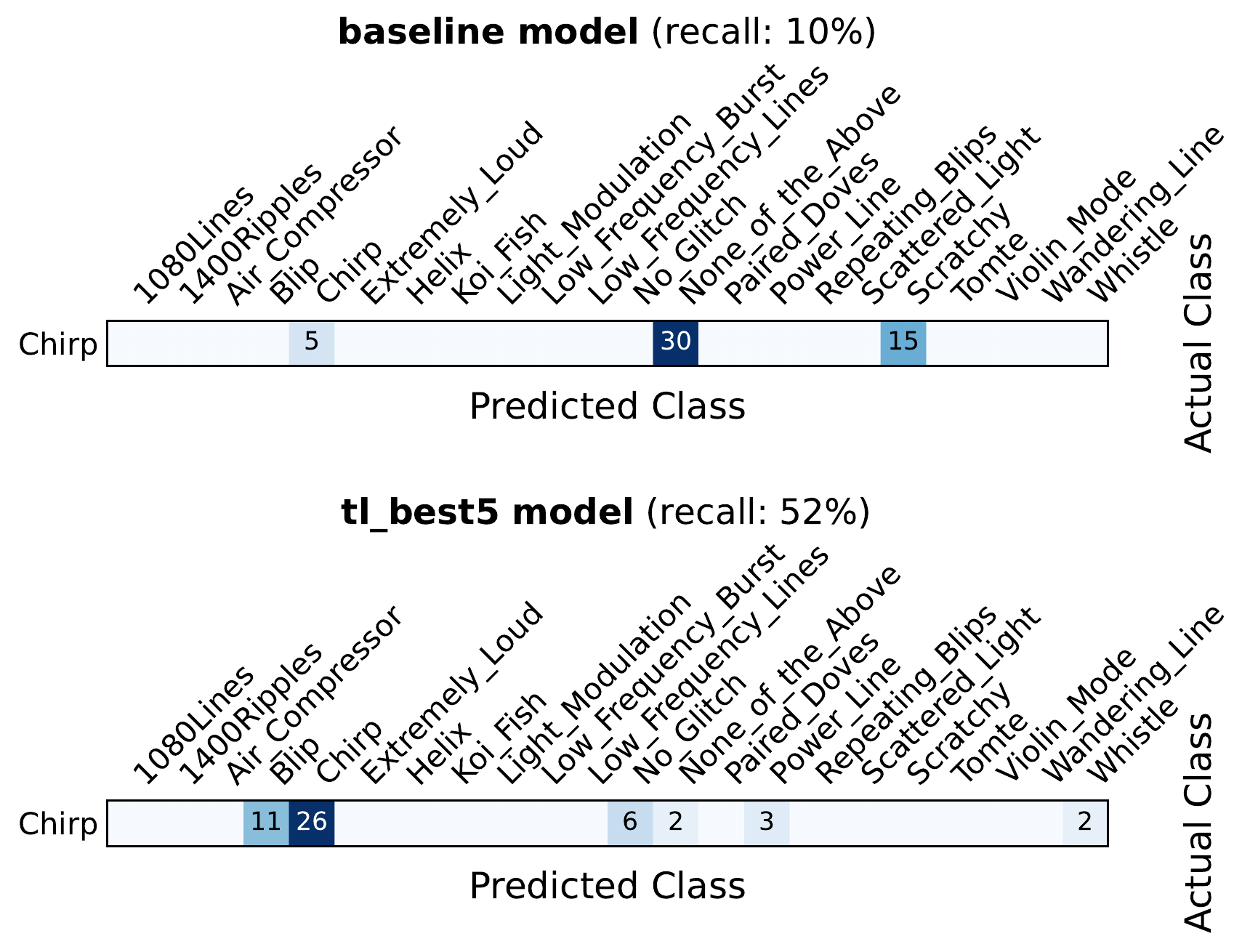}
        \caption{ \label{fig: o3gws lmatrices}}
    \end{subfigure}
    \begin{subfigure}[b]{0.495\linewidth}
        \centering
        \includegraphics[width=\linewidth]{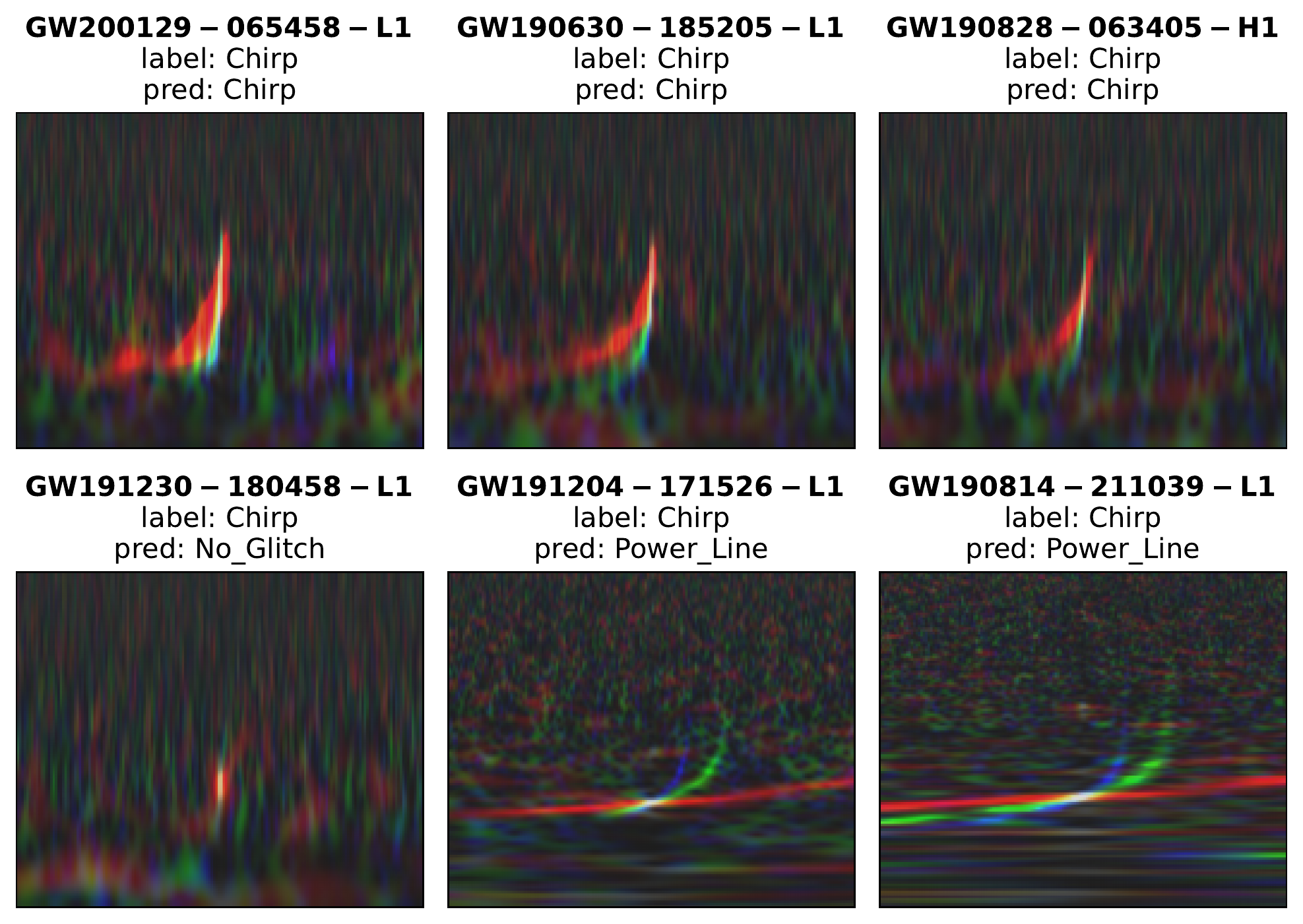}
        \caption{ \label{fig: o3gws samples}}
    \end{subfigure}
    \caption{Results of the predictions of the supervised models for O3 GW images. Left panel (a): Confusion matrices with a single row each corresponding to the \texttt{Chirp} class. Right panel (b): examples of the tl\_best5 model predictions. 
    \label{fig:losses}}
\end{figure*}

The views with 0.5, 2.0 and 4.0 seconds of duration were combined taking advantage of the RGB color channels, as done previously, to obtain the \texttt{encoded134} view.
The resulting images were passed through the baseline and \texttt{tl\_best5} model. The results are shown in panel (a) of Figure~\ref{fig:losses}. 
The performance of the two supervised models tested with real GW signals from the O3 run is noticeably different, despite both having perfect performance in the \texttt{chirp} class of the test dataset. On the one hand, the baseline model is very dependent on the image creation pipeline. For this reason, an extra alignment step for the colour channels was necessary to improve its performance. Even for the aligned images, the model achieves a mere 10\% recall, assigning most images to the \texttt{None\_of\_the\_Above} class. On the other hand, the transfer learning model achieves a 52\% recall for the aligned images, correctly identifying 26 out of the 50 GW samples. 
Moreover, the model is much more robust to samples which are different from the ones used to train it (i.e.~with different background noise conditions), being practically unaffected by the colour channel misalignment. One possible explanation for the better performance and behaviour with samples slightly out of distribution could be the benefit of pre-trained features, which were reported in \cite{Yosinski2014a} to produce a boost in generalization even after fine-tuning.

Panel (b) of Figure~\ref{fig:losses} displays an illustrative sample of the O3 GW signals used to evaluate the model, as well as the respective tl\_best5 model predictions. As expected, we found that the model performs well with chirps similar to those present in the Gravity Spy dataset (top row), typically from BBH mergers, but has more difficulties with different types of chirps (bottom row) which are either much longer (in the case of BNS coalescences) or much narrower (in the case of high-mass BBH mergers) than the chirps present in the training dataset.

\begin{figure*}[!htp]
    \begin{subfigure}[b]{0.495\linewidth}
        \centering
        \includegraphics[scale=0.5]{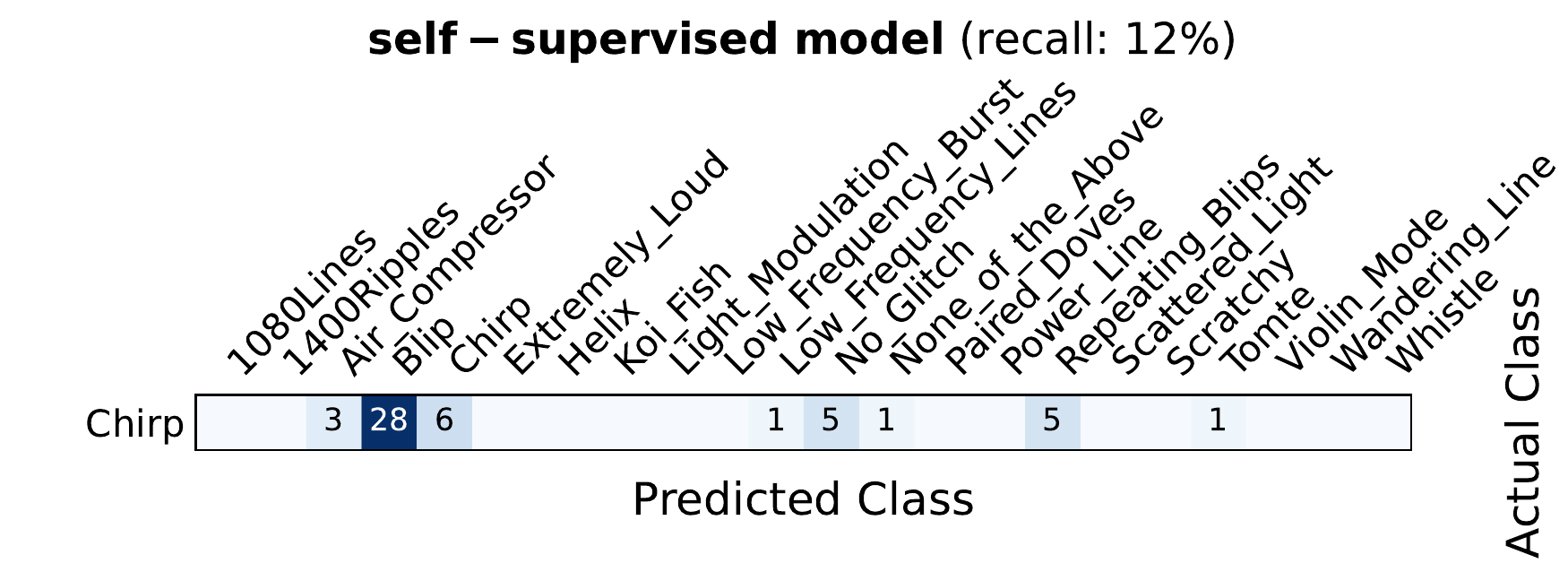}
        \caption{ \label{fig: ssl_o3gws}}
    \end{subfigure}
    \begin{subfigure}[b]{0.495\linewidth}
        \centering
        \includegraphics[scale=0.22]{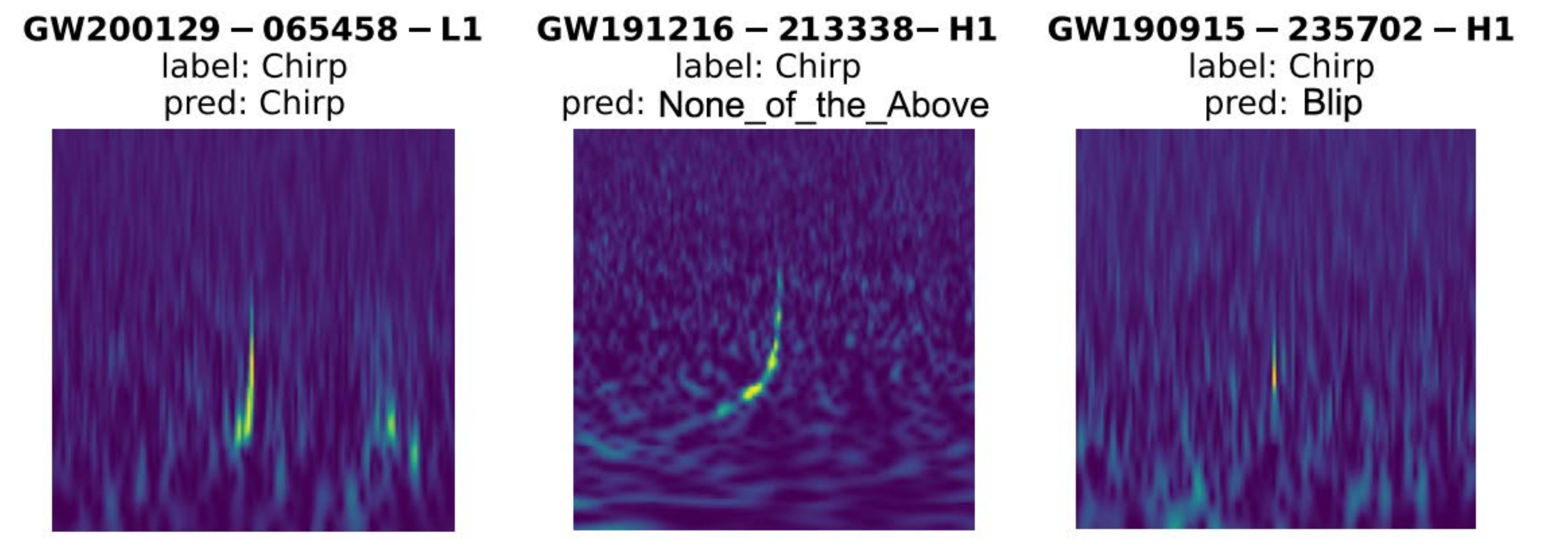}
        \caption{ \label{fig: ssl_o3gws}}
    \end{subfigure}
    \caption{Results of the predictions of the self-supervised model for O3 GW images. Left panel (a): A confusion matrix with only one row corresponding to the \texttt{Chirp} class. The majority of samples are classified as \texttt{Blips}. Right panel (b): examples of the model predictions together with its input images.}
    \label{fig:losses2}
\end{figure*}

Finally we discuss the results for the self-supervised model, applied to the 4-second view images. The results are plotted in Figure~\ref{fig:losses2}. Panel (a) of this figure indicates that most of the classified images are wrongly predicted as \texttt{Blips}. This is likely due to the fact that \texttt{Chirps} and \texttt{Blips} in the 4-second window images are quite similar to one another, as observed in panel (b) of Figure~\ref{fig:losses2}. In addition, the relatively small number of samples of \texttt{Chirps} in the dataset used for training might contribute to justify the very low GW detection rate. Nevertheless, the model has a similar recall value than the baseline supervised model. However, the value is significantly smaller than that achieved by the tl\_best5 model. 

Further studies to improve the performance of both the supervised and the self-supervised models by using an appropriate training on O3 data will be conducted in the future. We note, however, that no reliably labelled O3 dataset is yet available in the Gravity Spy project. The labels from the O3 dataset reported in~\cite{Glanzer:2023} were automatically obtained using a model trained mostly using O1/O2. Until this dataset becomes available, the next step for self-supervised models will be to use encoded views similar to the supervised models. This could allow us to understand if self-supervised methods are worth pursuing.

\section{Conclusion}
\label{section: conclusion}

The classification and mitigation of transient sources of noise, or ``glitches", in GW data streams is a crucial task in detector characterization and GW data analysis. In this paper we have investigated the use of Convolutional Neural Networks to classify glitches included in the Gravity Spy dataset, corresponding to the O1 and O2 data-taking periods of the Advanced LIGO and Advanced Virgo detectors. We have used both supervised and self-supervised deep learning.
First, we have trained models using a supervised learning approach, both trained from scratch and using transfer learning and hence fine-tuning already pre-trained models in the Gravity Spy dataset. The comparison with the best baseline model when using transfer learning shows a moderate increase of the F1 score (our metric of choice to compare different model configurations)  from 98.07\% to 98.21\%. However, the transfer learning configuration results in more stable performances across multiple training trials. In the second part of this work we have assessed the use of self-supervised training. In this case models have been pre-trained with pseudo-labels corresponding to image transformations applied to the original images, and then fine-tuned with the original labels. 

The results obtained from our best models are very close to the state-of-the-art results reported in the literature using the Gravity Spy dataset. Our best baseline and transfer learning supervised models reach accuracies higher than the merged view CNNs reported in~\cite{Bahaadini2017,Bahaadini2018}. Moreover, the baseline model's performance is also better than the hard fusion ensemble of CNNs~\cite{Bahaadini2018} and only slightly worse than the fine-tuned ResNet50 results from~\cite{George2018}, despite the fact we use a significantly simpler ResNet18 architecture trained from scratch. On the other hand, the scores of our self-supervised models show slightly lower values with respect to supervised approaches, yielding an F1 score of 94.15\%. This value, however, is close to the best performance achieved in Ref.~\cite{Bahaadini2018} using a single view.

In the last part of this study we have tested the models using actual GW signals from LIGO-Virgo's O3 run. We have found that despite the models having been trained using data from previous runs (O1 and O2), they show good performance, in particular the supervised model with transfer learning. 
When using transfer learning the scores improve in a significant way without the need for any training on actual GW signals (other than the less than 50 chirp examples from hardware injections present in the Gravity Spy dataset). This finding motivates the use of transfer learning not only for glitch classification but also for GW classification, as the model flexibility and undemanding generalization might help detect signals slightly different from the ones used during training. Transfer learning may thus provide a better coverage for glitch classification and GW detection with our ever-changing GW detection network.

To end, we note that an experiment worth performing elsewhere would be to fine-tune the transfer-learning model on O3 data. For this, we would need to classify O3 glitches for each class (including the new ones as e.g.~\texttt{Fast Scattering}~\cite{Soni:2021}) by hand, both for training and for testing.
The task of producing a reliable O3 glitch dataset (like there is for O1 and O2) would probably take significant effort, but perhaps the outputs of our models, the Gravity Spy team's model, and the human labelling from the Gravity Spy citizen science project can help.

\begin{acknowledgments}

%
We thank Osvaldo Freitas and Solange Nunes for fruitful discussions during the course of this work. 
We also thank Christopher Berry, Jools Clarke, Tom Dooney, Melissa López, Jade Powell, Max Razzano, and Agata Trovato for useful comments.
AO is supported by the FCT project CERN/FIS-PAR/0029/2019.
JCB is supported by a fellowship from ``la Caixa'' Foundation (ID 100010434) and from the European Union’s Horizon 2020 research and innovation programme under the Marie Skłodowska-Curie grant agreement No 847648 (fellowship code LCF/BQ/PI20/11760016). JCB is also supported by the research grant PID2020-118635GB-I00 from the Spain-Ministerio de Ciencia e Innovaci\'{o}n.
ATF and JAF are supported by the Spanish Agencia Estatal de Investigaci\'on (Grant  PID2021-125485NB-C21) funded by MCIN/AEI/10.13039/501100011033 and ERDF A way of making Europe). Further support is provided by the EU's Horizon 2020 research and innovation (RISE) programme H2020-MSCA-RISE-2017 (FunFiCO-777740) and  by  the  European Horizon  Europe  staff  exchange  (SE)  programme HORIZON-MSCA-2021-SE-01 (NewFunFiCO-10108625).
This material is based upon work supported by NSF's LIGO Laboratory which is a major facility fully funded by the National Science Foundation.

\end{acknowledgments}

\bibliographystyle{apsrev}
\bibliography{references}

\end{document}